\documentclass[review]{elsarticle}

\usepackage{hyperref}

\usepackage{graphicx}
\usepackage{amsmath}
\usepackage{amssymb}
\usepackage{mathtools}
\usepackage[version=4]{mhchem}
\usepackage{physics}
\usepackage{siunitx}
\usepackage{enumerate}
\usepackage{here}

\journal{Nuclear Instrument and Method A}

\begin{document}

\begin{frontmatter}

\title{Design and performance of a scintillation tracker\\for track matching in nuclear\nobreakdash-emulsion\nobreakdash-based\\neutrino interaction measurement}

\author[kyotoaddress]{T.~Odagawa\corref{mycorrespondingauthor}}
\cortext[mycorrespondingauthor]{Corresponding author}
\ead{odagawa.takahiro.57w@st.kyoto-u.ac.jp}

\author[nagoyaaddress]{T.~Fukuda}

\author[kyotoaddress]{A.~Hiramoto\fnref{footnote:hiramoto}}
\fntext[footnote:hiramoto]{Now at Okayama University}

\author[nagoyaaddress]{H.~Kawahara}

\author[kyotoaddress]{T.~Kikawa}

\author[yokohamaaddress]{A.~Minamino}

\author[kyotoaddress]{T.~Nakaya}

\author[nagoyaaddress]{O.~Sato}

\author[nagoyaaddress]{Y.~Suzuki}

\author[kyotoaddress]{K.~Yasutome}

\address[kyotoaddress]{Kyoto University, Kyoto 606-8502, Japan}
\address[nagoyaaddress]{Nagoya University, Nagoya 464-8602, Japan}
\address[yokohamaaddress]{Yokohama National University, Yokohama 240-8501, Japan}

\begin{abstract}

Precise measurement of neutrino\nobreakdash-nucleus interactions with an accelerator neutrino beam is highly important for current and future neutrino oscillation experiments.
To measure muon-neutrino charged-current interactions with nuclear-emulsion-based hybrid detector, muon track matching among the detectors are essential.
We describe the design and performance of a newly developed scintillation tracker for the muon track matching in the neutrino\nobreakdash-nucleus interaction measurement with nuclear emulsion detectors.
The muon tracks are reconstructed using the scintillation tracker and another detector called Baby MIND, then, they are matched with the tracks in nuclear emulsion detectors.

The scintillation tracker consists of four layers of horizontally and vertically aligned scintillator bars, covering an area of $\SI{1}{\m} \times \SI{1}{\m}$.
In the layer, \SI{24}{\mm}-wide plastic scintillator bars are specially arranged with deliberate gaps between each other.
By recognizing the hit pattern of the four layers, a precise positional resolution of \SI{2.5}{\mm} is achieved while keeping the number of readout channels as small as 256.
The efficiency of the track matching is evaluated to be more than 97\% for forward-going muons, and the positional and angular resolutions of the scintillation tracker are \SI{2.5}{\mm} and 20--\SI{40}{\milli\radian} respectively.
The results demonstrate the usefulness of the design of the scintillation tracker for the muon track matching in the nuclear\nobreakdash-emulsion\nobreakdash-based neutrino\nobreakdash-nucleus interaction measurements.

\end{abstract}


\end{frontmatter}

\section{Introduction\label{sec:introduction}}

Precise measurement of the neutrino oscillation parameters is one of the main topics in high energy physics.
In the current and future long-baseline neutrino oscillation experiments~\cite{T2K:2021xwb, NOvA:2021nfi, Abe:2018uyc, Acciarri:2015uup}, one of the major sources of the systematic uncertainties is the uncertainty of neutrino\nobreakdash-nucleus interaction models.
This uncertainty of the models comes from complicated nuclear effects.
To better understand such effects, a measurement of hadrons from the neutrino\nobreakdash-nucleus interactions in the sub- and multi-GeV energy region is required with a low momentum threshold and large angle acceptance.
In this energy region, protons from the interactions typically have momentum from a few to several hundred $\si{\MeV}/c$.

Neutrino Interaction research with Nuclear emulsion and J\nobreakdash-PARC Accelerator (NINJA) is an experiment aiming to measure neutrino\nobreakdash-nucleus, especially neutrino\nobreakdash-water, interactions using a high\nobreakdash-intensity neutrino beam produced in Japan Proton Accelerator Research Complex (J\nobreakdash-PARC), and nuclear emulsion films.
Since 2014, the NINJA experiment has carried out a series of pilot runs~\cite{Fukuda:2017clt, Hiramoto:2020gup, Oshima:2020ozn}.
Nuclear emulsion films have sub-\si{\um} positional resolution and $\order{\si{\milli\radian}}$ angular resolution thanks to their fine granularity.
It allows us to detect very short tracks of low\nobreakdash-momentum charged particles, especially protons down to $\sim \SI{200}{\MeV}/c$, which are of interest to evaluate 2\nobreakdash-particle 2\nobreakdash-hole (2p2h) neutrino interaction models~\cite{Nieves:2011pp, Martini:2009uj} where one charged lepton and two nucleons are emitted in the final state.
The results from the NINJA experiment will improve our understanding and constrain neutrino interaction models in the sub- and multi-GeV neutrino energy region.

The NINJA experiment conducted its first physics run (J-PARC E71a) from November 2019 to February 2020.
In this run, a \SI{250}{\kg} (including \SI{75}{\kg} water, \SI{130}{\kg} iron, \SI{15}{\kg} plastic, and \SI{30}{\kg} emulsion) nuclear emulsion detector was exposed to the T2K neutrino beam produced at J\nobreakdash-PARC.
The total number of protons on target (POT) was $4.7 \times 10^{20}$ and the mean neutrino energy was \SI{0.89}{\GeV}.
A precise measurement of neutrino\nobreakdash-water differential cross sections will be achieved with a target mass 19 times larger than the one in the previous run in 2017--18~\cite{Hiramoto:2020gup}.

To study muon neutrino charged\nobreakdash-current ($\nu_\mu$ CC) interactions, we placed not only nuclear emulsion detectors but also a muon range detector (MRD) which distinguishes between muons and charged pions from the interactions.
The nuclear emulsion films can measure position and angle very precisely while they do not have any timing information.
On the other hand, the MRD has an $\order{\si{\ns}}$ timing resolution but the positional and angular resolutions are worse than those of the nuclear emulsion films.
Thus, to match the muon tracks between the nuclear emulsion detectors and the MRD, other detectors which have both good positional and angular resolutions, and timing information were installed between them.

In the previous runs of the NINJA experiment, an emulsion shifter~\cite{Oshima:2020ozn, Yamada:2017qeg} or a scintillating fiber tracker~\cite{Hiramoto:2020gup} was used for the muon track matching and covered an area around $\SI{30}{\cm} \times \SI{30}{\cm}$.
The emulsion shifter was originally developed for cosmic-ray electron balloon experiments~\cite{Kodama:2006zz} and used in the GRAINE experiment~\cite{Takahashi:2016xsf, Takahashi:2018nkq}.
It provides precise information of position and angle of the nuclear emulsion films while the timing resolution was 5--\SI{50}{\s} in the previous runs.
The scintillating fiber tracker consisted of 1,024 scintillating fibers with a cross section of 1\,mm-square.
It has a \SI{1}{\ns} timing resolution and a few hundred \si{\um} positional resolution but the number of readout channels were more than 500 and the enlargement is not easy\footnote{To cover an area of $\SI{1}{\m} \times \SI{1}{\m}$, the number of fibers needed will be more than 3,000.}.
In the physics run, such detectors must cover an area of $\SI{1}{\m} \times \SI{1}{\m}$, which is around ten times larger than that in the previous runs, and the emulsion shifter and the scintillator detector are both used for this purpose.
The combination reduces the required positional resolution for the scintillator detector; however, it is still difficult to enlarge the coverage area while maintaining the number of readout channels similar to that of the previous one.
To realize both the large coverage area and the good positional resolution without increasing the number of readout channels, we developed a new scintillation tracker with a special arrangement of plastic scintillator bars.
By recognizing the hit pattern, the scintillation tracker achieves a few mm positional resolution while the width of the bar is \SI{24}{\mm}.
In addition, it has a $\SI{1}{\m} \times \SI{1}{\m}$ coverage area without increasing the number of readout channels nor insensitive areas.

In this paper, the design and performance of the scintillation tracker for the NINJA physics run are described.
In Section~\ref{sec:setup}, requirements for the scintillation tracker are described after the conceptual setup of the NINJA detectors.
The design and components of the scintillation tracker are described in Section~\ref{sec:design}.
The operation of the scintillation tracker during the beam exposure is summarized in Section~\ref{sec:operation}.
The reconstruction of muon tracks for the $\nu_\mu$ CC interaction analysis are described in Section~\ref{sec:matching}.
The hit-track matching efficiency, and positional and angular resolutions of the scintillation tracker are shown in Section~\ref{sec:performance}.
Finally, Section~\ref{sec:summary} summarizes the paper.

\section{NINJA detector setup and requirements for the scintillation tracker\label{sec:setup}}

\subsection{Detectors\label{ssec:setup:detectors}}

\begin{figure}[h]
\centering
\includegraphics[width = 0.8 \textwidth]{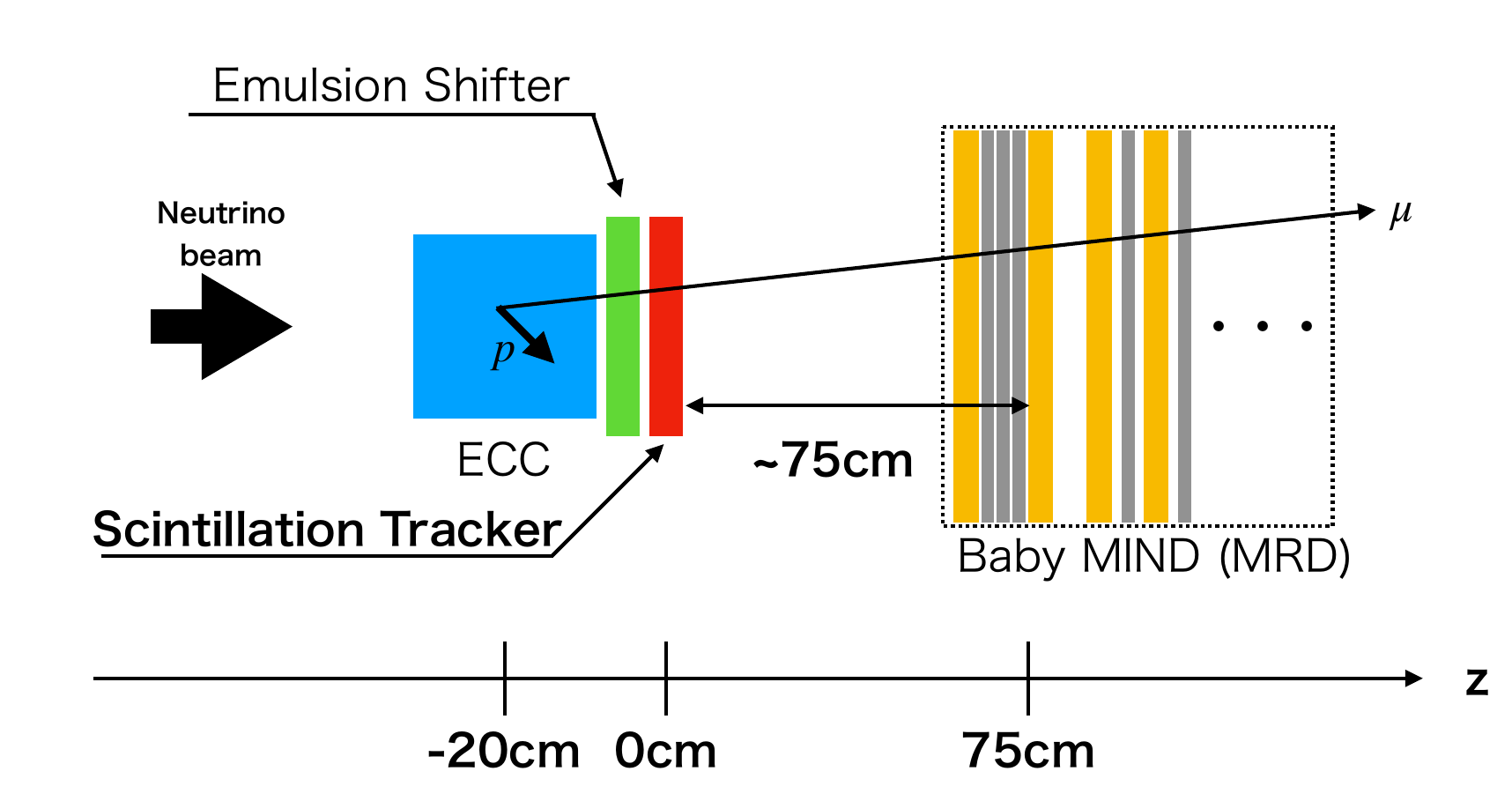}
\caption{Detector setup of the NINJA physics run. The ECC was installed in the most upstream of the neutrino beam and detected charged particle tracks from the neutrino interaction vertices. A muon was detected in Baby MIND in the most downstream. Baby MIND consists of detector modules (orange) and magnet modules (gray). An emulsion shifter and a scintillation tracker were installed between them to connect muon tracks. The $z$ direction is defined along the beam and the location of the scintillation tracker is set as $z = \SI{0}{\cm}$.\label{fig:setup:detector_setup}}
\end{figure}

In the NINJA physics run, four types of detectors were installed and operated in the J\nobreakdash-PARC Neutrino Monitor (NM) building.
The detectors were exposed to a high\nobreakdash-intensity neutrino beam dominated by muon neutrinos~\cite{T2K:2012bge}.
A schematic of the detector setup is shown in Fig.~\ref{fig:setup:detector_setup}.
Here, the $z$ direction is defined along the neutrino beam and the location of the scintillation tracker is set as $z = \SI{0}{\cm}$.

The main target detector, Emulsion Cloud Chamber (ECC), has an alternating structure consisting of thin target material layers and nuclear emulsion films.
This structure allows us to achieve a low momentum threshold for charged particles from the neutrino interactions in any target material.
In the NINJA physics run, water and iron layers were used as the target material.
Water was the main target.
Iron, apart from being the target layer, was used as the support of the nuclear emulsion films and for momentum measurement using range or multiple Coulomb scatterings~\cite{Hiramoto:2020gup}.
Each ECC has a dimension of $\sim \SI{30}{\cm} \times \SI{30}{\cm} \times \SI{30}{\cm}$.
In the NINJA physics run, nine ECCs were arranged in a $3 \times 3$ array on a plane perpendicular to the beam direction; thus, the total coverage area was approximately $\SI{1}{\m} \times \SI{1}{\m}$.

The total thicknesses of water and iron layers in each ECC along the beam direction are \SI{133}{\mm} and \SI{35}{\mm}, respectively.
Thus, the total thickness of ECC is $\sim 2.5$ radiation length units.
Low\nobreakdash-momentum protons stop inside the ECC volume, while almost all muons and charged pions from the neutrino interactions penetrate the ECC.
To distinguish the muons from the charged pions for studying $\nu_\mu$ CC interactions, the ECCs were placed upstream of an MRD called Baby MIND (prototype of Magnetized Iron Neutrino Detector), one of the T2K near detectors~\cite{Parsa:2020nyi}.

A muon from $\nu_\mu$ CC interaction is detected and identified by Baby MIND.
Baby MIND consists of 18 detector modules and 33 magnet modules.
The detector module consists of plastic scintillator bars and detects hits of a muon.
The magnet module is a \SI{3}{\cm}-thick magnetized iron plane.
Its thickness is sufficient to stop muons of energy $\sim \SI{1}{\GeV}/c$ inside the detector.
The momentum and charge of the muon can be measured by the range and curvature of the trajectory.

The nuclear emulsion film cannot provide timing information.
The films record all the tracks after production and the tracks are developed at once after the beam exposure.
Therefore, many cosmic-muon tracks are recorded as a background.
On the other hand, Baby MIND has beam-timing information: Baby MIND data acquisition (DAQ) is triggered with a beam timing signal and the time resolution of readout electronics is \SI{2.5}{\ns}.
To identify muons from the neutrino interactions in the ECC, the tracks in the ECCs need to be connected to corresponding tracks in Baby MIND.
However, Baby MIND has a positional resolution of $\order{\si{\cm}}$.
It is insufficient to select a correct track to be connected from all the tracks accumulated in the nuclear emulsion films whose density is $\order{10^4}/\si{\cm^2}$.
To connect the tracks one by one, other detectors with both good positional and angular resolutions, and timing information are needed.
Such detectors are called ``timestamp detectors.''
As mentioned in Section~\ref{sec:introduction}, two types of timestamp detectors were installed and operated in the NINJA physics run.
One is an emulsion shifter and the other is a scintillation tracker.

The emulsion shifter has several nuclear emulsion films attached on walls shifted at different time intervals.
After the films are developed, the tracks in the films are connected to each other with certain shifts and these shifts give timing information of the connected tracks.
The newly developed emulsion shifter for the NINJA physics run consists of emulsion films attached on two moving walls and one fixed wall.
One moving wall shifts by \SI{2}{\mm} every four hour and the other also does by \SI{2}{\mm} every four day.
The positional and angular information are obtained with $\sim \SI{5}{\um}$ and $\SI{4}{\milli\radian}$ resolutions of the nuclear emulsion films, and 4-hour timing information is provided by the shift value.
Using the positional and angular information, the tracks are extrapolated and connected to the ones in the nuclear emulsion films of the ECC.
Details of the emulsion shifter will be reported in another paper.

The scintillation tracker is used to connect tracks between the emulsion shifter and Baby MIND.
Although the emulsion shifter has excellent positional and angular resolutions, and 4-hour timing information, during the whole period, a large number of cosmic-muon tracks are accumulated.
To select a muon candidate in the beam timing, another timestamp detector whose DAQ is triggered with the beam timing signal is necessary.
The DAQ of the scintillation tracker is triggered with the same beam timing signal as that of Baby MIND.
Therefore, the hit-track matching between the scintillation tracker and Baby MIND is almost free from the cosmic background.
The reconstructed position and angle are more precise compared to those of the tracks reconstructed using only the information from Baby MIND.
Then, using the position and angle, and 4-hour timing information of the emulsion shifter, the tracks are connected between the emulsion shifter and the scintillation tracker.

\subsection{Requirements for the scintillation tracker\label{ssec:setup:requirement}}

Before the design and construction of the scintillation tracker, the requirements for the scintillation tracker was evaluated by rough estimations and data taken in pilot experiments.
Here, it should be noted that the requirements were conservatively estimated because the reconstruction methods of the position and angle in Baby MIND were not matured at that time.
The actual performances of the positional and angular resolutions of the reconstructed tracks are discussed in Section~\ref{sec:performance}.

\subsubsection{Positional resolution\label{sssec:setup:requirement:posres}}

As described in Section~\ref{ssec:setup:detectors}, good positional and angular resolutions are required for the scintillation tracker.
First, the angle of a muon track, $\theta_{x(y)}$, is calculated using reconstructed positions in Baby MIND and the scintillation tracker as
\begin{equation}
\tan\theta_{x(y)} = \frac{x(y)_{\mathrm{BM}} - x(y)_{\mathrm{ST}}}{d},
\label{eq:setup:thetarec}
\end{equation}
where $x(y)_{\mathrm{BM}}$ and $x(y)_{\mathrm{ST}}$ are horizontal (vertical) positions reconstructed by Baby MIND and the scintillation tracker, respectively, and $d \sim \SI{75}{\cm}$ is the distance between Baby MIND and the scintillation tracker. 

In addition, \SI{50}{\cm}-thickness water-target detector, WAGASCI, is placed between the scintillation tracker and Baby MIND, which is not shown in Fig.~\ref{fig:setup:detector_setup}.
The positional variation by the scattering by the water should be also taken into account.
Considering Eq.~\eqref{eq:setup:thetarec} and the effect from the scattering in WAGASCI, the angle resolution can be written as
\begin{equation}
\sigma_{\tan\theta_{x(y)}} = \frac{\sqrt{\sigma_{\mathrm{BM}, x(y)}^2 + \sigma_{\mathrm{ST},x(y)}^2 + \sigma_{\mathrm{MCS}}^2}}{d},
\label{eq:setup:thetarecres}
\end{equation}
where $\sigma_{\mathrm{BM}, x(y)}$ and $\sigma_{\mathrm{ST}, x(y)}$ are horizontal (vertical) positional resolutions of Baby MIND and the scintillation tracker, respectively, and $\sigma_{\mathrm{MCS}}$ is the positional variation by the scattering in WAGASCI.
For instance, $\sigma_{\mathrm{MCS}} \simeq \SI{16}{\mm}$ for $\SI{1}{\GeV}/c$ muons.

In the horizontal direction, the positional resolution of Baby MIND is much worse than that of the scintillation tracker because the effective width of one horizontal Baby MIND scintillator bar, $w_{\mathrm{BM}, x}$, is around \SI{21}{\cm} while the resolution of the scintillation tracker will be \SI{3}{\mm} as described later.
Besides, the scattering in WAGASCI is also negligible.
Therefore, Eq.~\eqref{eq:setup:thetarecres} can be approximated as
\begin{equation}
\begin{split}
    \sigma_{\tan\theta_{x}} &\sim \frac{\sigma_{\mathrm{BM}, x}}{d} \\ 
&= \frac{w_{\mathrm{BM}, x}}{\sqrt{12} d},
\label{eq:setup:thetarecresapprox_x}
\end{split}
\end{equation}
where $\sigma_{\mathrm{BM}, x}$ can be written as $\sigma_{\mathrm{BM}, x} = w_{\mathrm{BM}, x}/\sqrt{12}$ if we assume the hits in each scintillator bar are uniformly distributed.

In the vertical direction, on the other hand, the effective width of one Baby MIND scintillator bar, $w_{\mathrm{BM}, y}$, is \SI{1}{\cm}.
Thus, the positional resolutions of Baby MIND and the scintillation tracker will be comparable if we require that of the scintillation tracker to be \SI{3}{\mm} as described later.
The positional variation by the scatterings is much larger than the positional resolutions and dominant.
Thus, Eq.~\eqref{eq:setup:thetarecres} can be approximated as
\begin{equation}
    \sigma_{\tan\theta_{y}} \sim \frac{\sigma_{\mathrm{MCS}}}{d}.
\label{eq:setup:thetarecresapprox_y}
\end{equation}
Using Eqs.~\eqref{eq:setup:thetarecresapprox_x} and \eqref{eq:setup:thetarecresapprox_y}, the angular resolutions of the reconstructed tracks from the scintillation tracker and Baby MIND are \SI{81}{\milli\radian} and \SI{20}{\milli\radian} for the horizontal and vertical directions, respectively.

\begin{figure}[H]
\centering
\includegraphics[width = 0.7\textwidth]{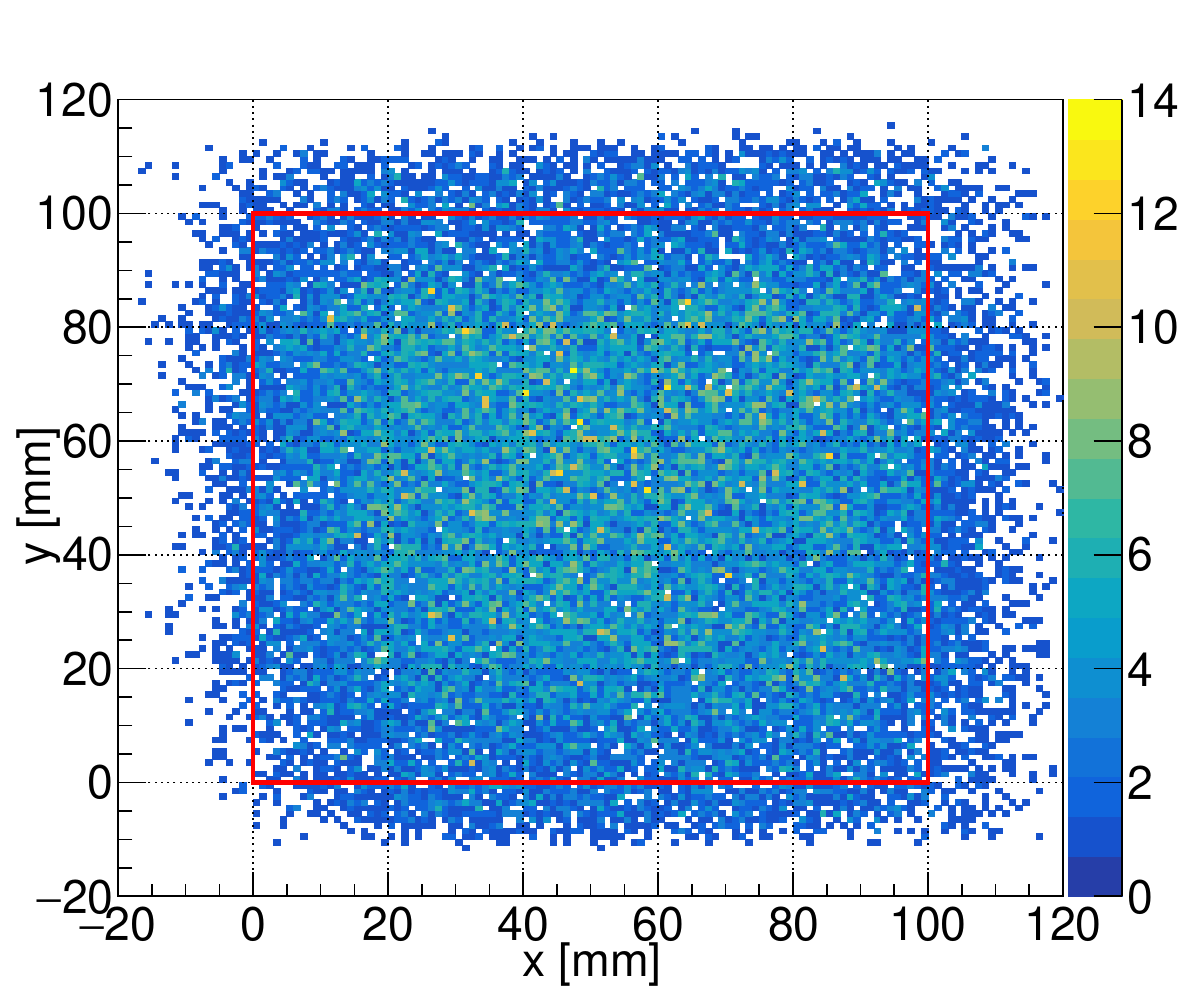}
\caption{Distribution of position of the tracks accumulated over a period of one month in the NINJA previous run. The tracks inside a red rectangle is used in the evaluation of the positional resolution.\label{fig:setup:run8_cs_pos}}
\end{figure}

\begin{figure}[H]
\centering
\includegraphics[width = 0.7\textwidth]{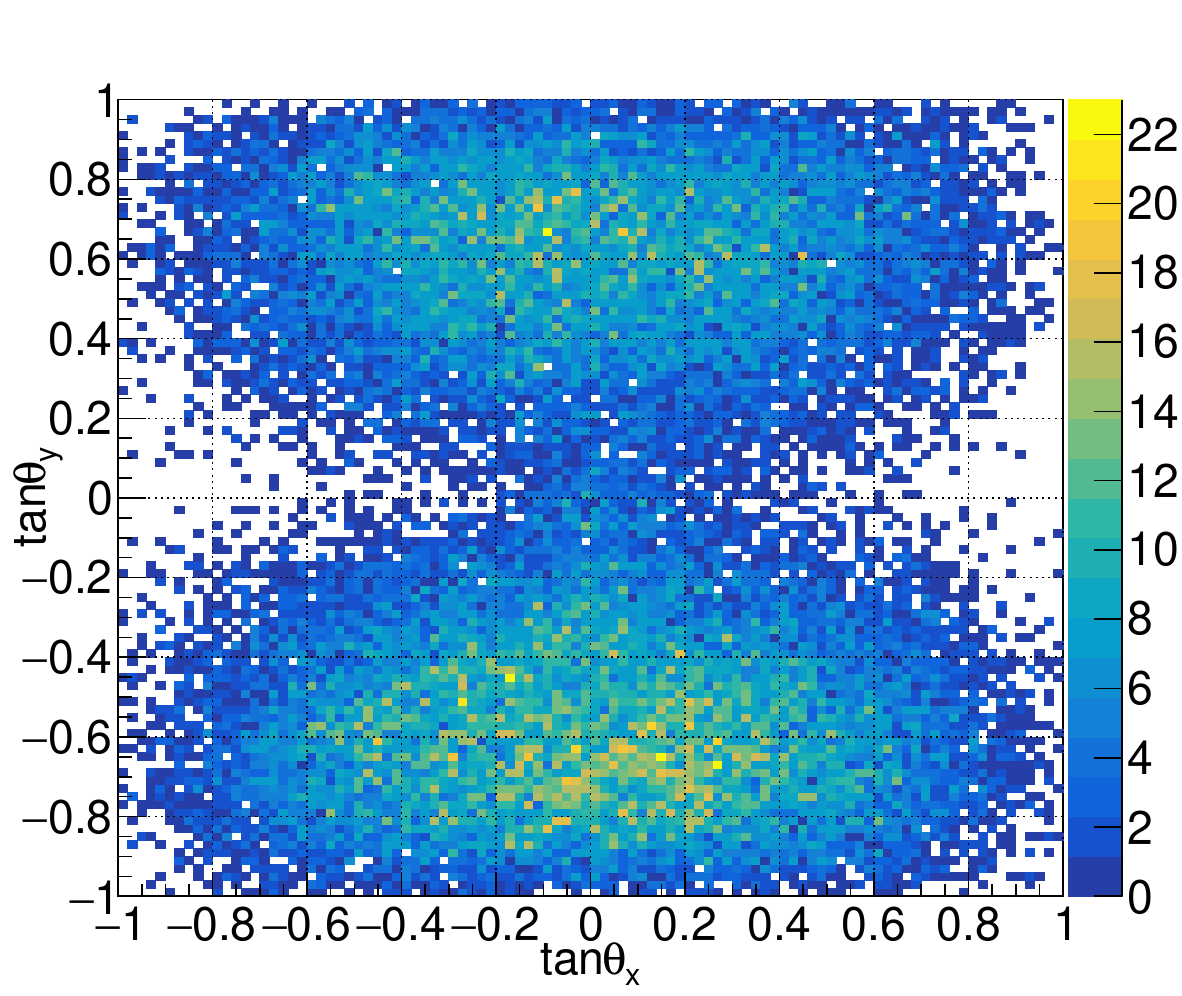}
\caption{Distribution of angle of the tracks accumulated over a period of one month in the NINJA previous run. The track density is highest around the point $(\tan\theta_x, \tan\theta_y) = (0, -0.6)$.\label{fig:setup:run8_cs_ang}}
\end{figure}

Next, using this angular resolution, the positional resolution requirement for the scintillation tracker is estimated as follows.
In this evaluation, track data from the previous run~\cite{Hiramoto:2020gup} are used.
Figures~\ref{fig:setup:run8_cs_pos} and \ref{fig:setup:run8_cs_ang} show the positional and angular distributions of tracks accumulated over a period of one month in an emulsion film.
The dominant component of the tracks is from cosmic muons.
Figure~\ref{fig:setup:run8_cs_pos} shows that the positional distribution is uniform except at the edge of the distribution.
The number of events in the edges is small due to the acceptance of the detectors.
To evaluate the positional resolution requirement for the scintillation tracker, tracks in an area with $x(y) = 0\text{--}\SI{100}{\mm}$ are used.
The area is shown as a red rectangle in Fig.~\ref{fig:setup:run8_cs_pos}.
Figure~\ref{fig:setup:run8_cs_ang} shows a characteristic angular distribution due to the zenith angle dependence of the cosmic muons and the acceptance of the film scanning system, Hyper Track Selector~\cite{Yoshimoto:2017ufm}.
In the angular space, the track density is highest around the point $(\tan\theta_x, \tan\theta_y) = (0, -0.6)$.
The number of tracks within the $(\pm 2 \sigma_{\tan\theta_x}) \times (\pm 2 \sigma_{\tan\theta_y}) = (4 \times \SI{81}{\milli\radian}) \times (4 \times \SI{20}{\milli\radian})$ region around this point is 730.
Thus, when we require $2 \sigma_{\tan\theta_{x(y)}}$ separation to select a correct track to be connected, 730 tracks cannot be separated.
Therefore, the density of the tracks which should be distinguished using positional information is calculated as $730~\text{tracks} / (\SI{100}{\mm} \times \SI{100}{\mm}) = 7.3 \times 10^{-2}~\text{tracks} / \si{\mm^2}$.
In addition, owing to the 4-hour timing information from the emulsion shifter, the density becomes $7.3 \times 10^{-2}~\text{tracks} / \si{\mm^2} \times \SI{4}{\hour} / 1\,\mathrm{month} \sim 4.1 \times 10^{-4}~\text{tracks} / \si{\mm^2}$.
The number of tracks within $(\pm 2 \sigma_\mathrm{ST})^2$ region is $(4 \sigma_\mathrm{ST}) \times (4 \sigma_\mathrm{ST}) \times (4.1\times10^{-4}~\text{tracks} / \si{\mm^2}) = 6.5 \times 10^{-3}\,\mathrm{tracks} / \si{\mm^2} \times \sigma_{\mathrm{ST}}^2$, where $\sigma_\mathrm{ST} = \sigma_{\mathrm{ST},x} = \sigma_{\mathrm{ST},y}$.
If we accept 5\% mis-matching contamination between the emulsion shifter and the scintillation tracker, then the requirement for the positional resolution is $6.5 \times 10^{-3}\,\mathrm{tracks/mm^2} \times \sigma_{\mathrm{ST}}^2 \leq 0.05$, i.e. $\sigma_{\mathrm{ST}} \leq \SI{2.8}{\mm}$.
This estimation is conservative because the detectors are at a \SI{5}{\m} lower floor than the previous run.
Since the cosmic-muons will encounter more material and some will be stopped, less cosmic-muons will reach that depth.
Therefore, the actual cosmic-muon flux in the physics run will be smaller.

\subsubsection{Size\label{sssec:setup:requirement:size}}

The scintillation tracker must cover an area of $\SI{1}{\m} \times \SI{1}{\m}$.
When a muon loses its energy in the scintillation tracker, the reconstructed momentum is biased.
In addition, when it is scattered in the material of the scintillation tracker, the position and angle change from the ones predicted by the emulsion shifter.
Thus, the scintillation tracker also needs to be thin enough to minimize the energy loss and scattering of muons.
Besides, the NINJA detectors were surrounded by the WAGASCI detectors of the T2K experiment.
The WAGASCI in the T2K experiment has several detectors as shown in Fig.~\ref{fig:setup:detector_setup}; the upstream WAGASCI module, the Proton Module, the downstream WAGASCI module, the north Wall MRD, the south Wall MRD, and Baby MIND.
The total thickness of the ECCs, the emulsion shifter, and the scintillation tracker along the beam direction was limited to less than \SI{50}{\cm}.
Therefore, the thin dimension of the scintillation tracker is also important for securing space for the ECCs to keep the target mass.
In the NINJA physics run, the thickness of the scintillation tracker is required to be less than \SI{5}{\cm}.
\begin{figure}[h]
\centering
\includegraphics[width = 0.5 \textwidth]{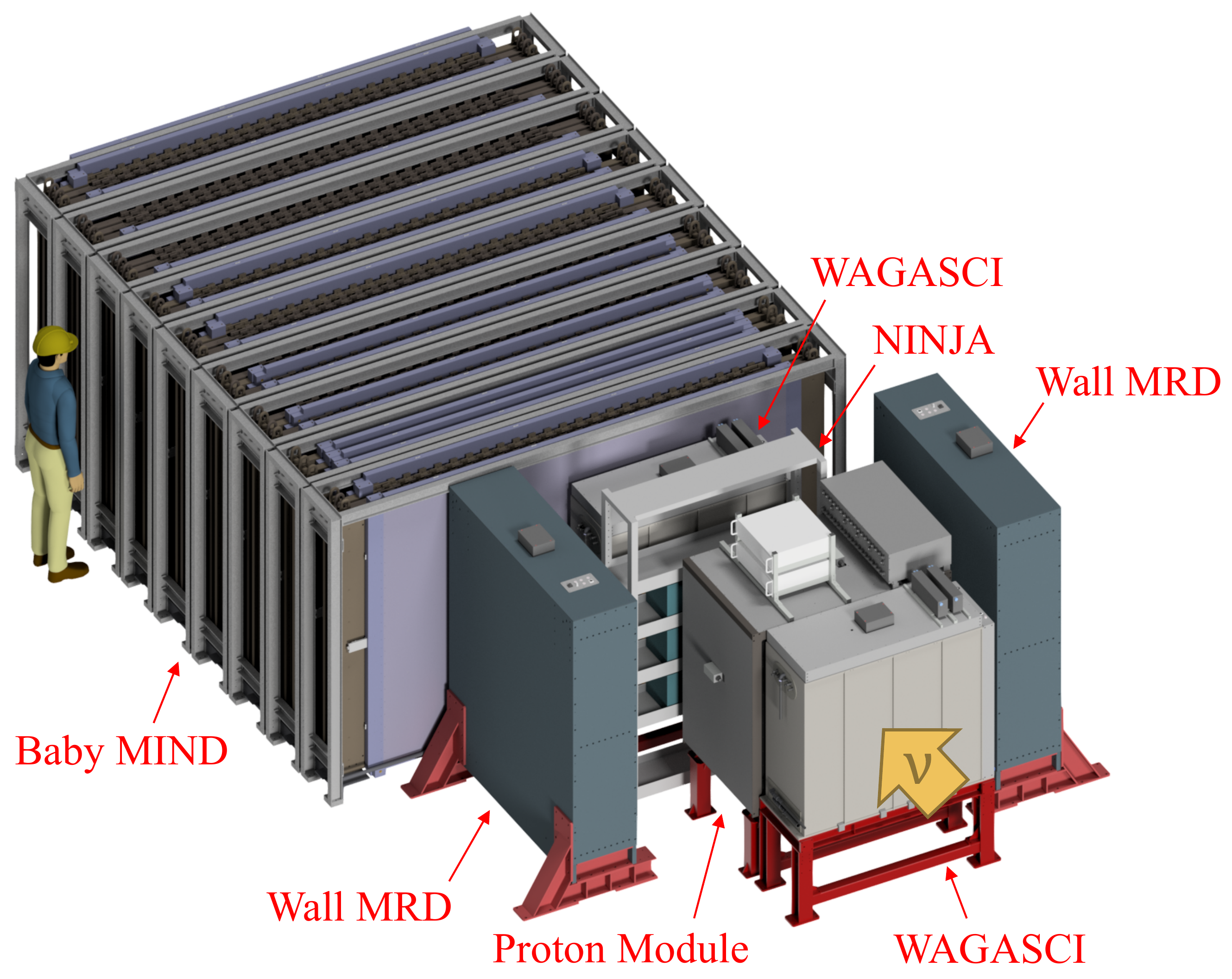}
\caption{Setup of the detectors in the J-PARC NM building. The NINJA detectors were surrounded by the WAGASCI detectors in the T2K experiment. The center positions of the detectors were not identical to maximize the muon angle acceptance of the WAGASCI detectors.\label{fig:setup:wagasci_babymind_w_ninja_w_note}}
\end{figure}

\subsubsection{Detection and matching efficiencies\label{sssec:setup:requirement:efficiency}}

Currently, the sensitivity of the NINJA experiment is limited by the statistics.
Thus, high detection and matching efficiencies of the detectors are essential to obtain the $\nu_\mu$ CC differential cross section by connecting as many muon tracks as possible.
In addition, to reduce the systematic uncertainties arising from the efficiencies in the $\nu_\mu$ CC interaction analysis, the efficiencies are also required to be as high as possible.
For example, when the efficiency is 95\%, the uncertainty of the number of detected muons will be at most 5\%, since it is proportional to the uncertainty of the efficiency.

The total muon detection efficiency is a product of the track detection efficiencies of the detectors and the track matching efficiencies between the detectors.
In the NINJA physics run, the tracks need to be matched between the ECC and the emulsion shifter, the emulsion shifter and the scintillation tracker, and the scintillation tracker and Baby MIND.
In this paper, we focus on the detection efficiency of the scintillation tracker and the matching efficiency between the scintillation tracker and Baby MIND.
In the analysis of the NINJA physics run, we set a target value of the efficiencies of the scintillation tracker at 95\%.
The other efficiencies will be described in another paper.
In the results from the NINJA pilot runs~\cite{Hiramoto:2020gup, Oshima:2020ozn}, the muon detection efficiency was dropped mainly due to inefficiency in matching between the emulsion detectors and the scintillator detectors which was around 90\%.
The improvement in this efficiency will be also reported in another paper.

\section{Design and components of the scintillation tracker\label{sec:design}}

To meet the requirements described in Section~\ref{ssec:setup:requirement}, a scintillation tracker with a new design was developed for the NINJA physics run.
The scintillation tracker provides positional information of the muon and it is matched with a track in Baby MIND.
The design has a special arrangement of relatively-shifted layers of thin plastic scintillator bars with deliberate gaps between the bars.
When a muon penetrates the gap, the position of the track can be constrained by the absence of hits.
The positional resolution obtained by this design is determined by the width of the gap and overlap between the bars; The positional resolution can be better than the width of the bar.

\subsection{Design\label{ssec:design:design}}

\begin{figure}[h]
\centering
\includegraphics[width = 0.8 \textwidth]{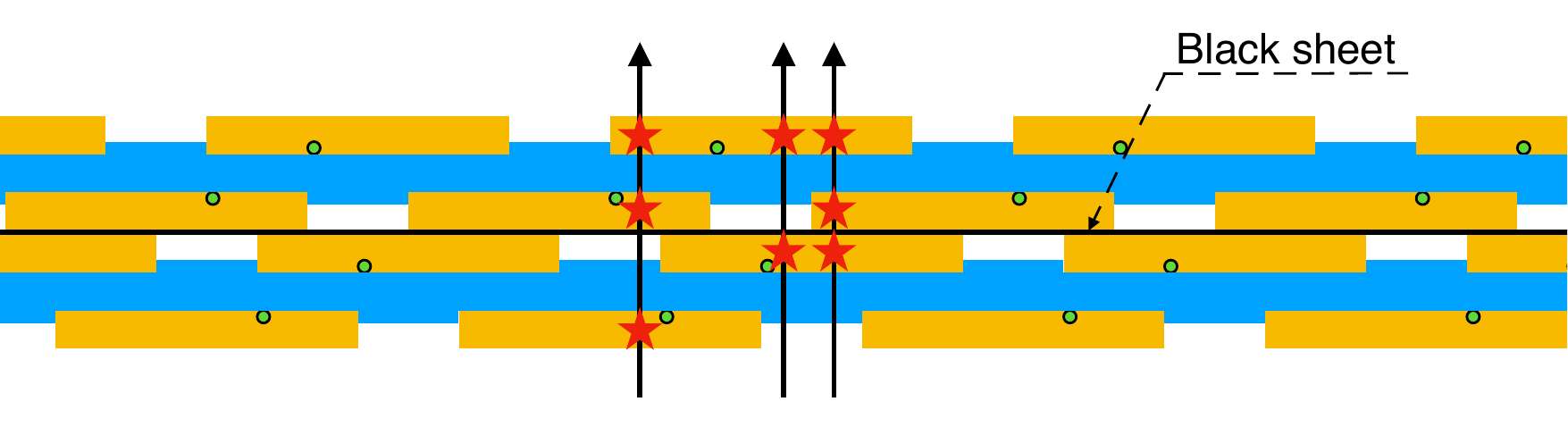}
\caption{Conceptual design of the scintillation tracker for the NINJA physics run. Each scintillator bar (orange) is aligned with a gap whose width is 1/3 of the scintillator bar. The scintillator bars are attached to and supported by the structure (blue). The black line between the second and third layers is a shading black sheet inserted to avoid optical cross-talk between the scintillators.\label{fig:design:tracker_design}}
\end{figure}

The scintillation tracker consists of two identical modules, each made of four layers.
One module is set with the scintillator bars horizontal and the other module is set with the bars vertical.
The surfaces of the both modules are perpendicular to the beam direction.
Figure~\ref{fig:design:tracker_design} shows the conceptual design of a module of the scintillation tracker.
The orange areas denote the scintillator active volume and the blue areas represent support structures made of polyvinyl chloride.
One layer is made of scintillator bars aligned with a gap equal to 1/3 of scintillator width.
The second layer is shifted by 2/3 of the scintillator width relative to the first layer, and the third and fourth layers are shifted by 1/6 of the scintillator width relative to the first and second layers, respectively.
Furthermore, in each module, a thin black sheet is inserted between the second and third layers to avoid optical cross-talk between the layers.
The positional resolution of the scintillation tracker improves without increasing the number of readout channels because it has a virtual segmentation of 1/6 of the width of the scintillator bar as shown in Fig.~\ref{fig:design:tracker_design} by getting the hit pattern in each module.

In addition to the good positional resolution, insensitive areas do not exist in this design.
The edges of the scintillator bar are generally insensitive.
However, viewed from the beam direction, the edges and gaps of the scintillator bars in one layer are fully covered by scintillator bars in the neighboring layer.
Thus the entire area of the scintillation tracker is sensitive and it achieves a detection efficiency of nearly 100\%.

If we assume that the muon track goes through in a direction perpendicular to the layers, and the probability distribution of the position in each virtual segmentation is uniform, the requirement for the scintillator width $w$ is $(w/6)/\sqrt{12} \leq \SI{2.8}{\mm}$ i.e. $w \leq \SI{6}{\cm}$.
In the real case, however, the muon tracks are not always perpendicular to the layers and the  reconstruction of the position is affected by the distortion or alignment of the scintillator bars.
Therefore, we conservatively require $w \leq \SI{3}{\cm}$.

\subsection{Components\label{ssec:design:component}}

The scintillation light from each bar is captured and propagated by a wavelength-shifting (WLS) fiber and the light is detected by a Multi-Pixel Photon Counter (MPPC).

\subsubsection{Plastic scintillator}

\begin{figure}[h]
\centering
\includegraphics[width = 0.9\textwidth]{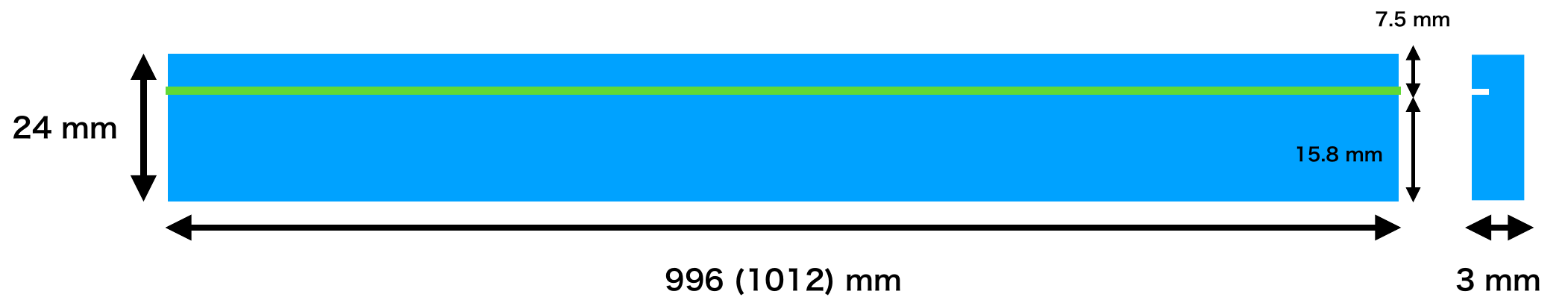}
\caption{The scintillator bar used in the tracker. A wavelength-shifting fiber (light green) is set on a groove. The thickness along the beam direction is \SI{3}{\mm} and the length and width are $\sim \SI{1}{\m}$ and \SI{24}{\mm}, respectively.\label{fig:design:scintillator}}
\end{figure}

The plastic scintillator bar is a main component of the scintillation tracker.
This scintillator bar is of the same type as the one used in the WAGASCI detector in the T2K experiment~\cite{T2K:2019dgm, Abe:2020iop}.
It is primarily composed of polystyrene and infused with PPO and POPOP.
The bar is coated by \ce{TiO2}-based white reflective paint in order to increase the light yield and optically separate each scintillator bar.
The scintillation tracker consists of 248 bars (124 bars for the horizontal and vertical modules respectively).
The thickness of the bar is \SI{3}{\mm} along the beam direction, and the length and width for the horizontal (vertical) module are 1012~(996)\,\si{\mm} and \SI{24}{\mm}, respectively.
The bar has a groove for a wavelength-shifting fiber and the depth of it is \SI{1.2}{\mm} on one side of the bar, as shown in Fig.~\ref{fig:design:scintillator}, and optical cement is applied not to lose the light yield.
The scintillator bar is thin enough that the energy loss for minimum ionizing particles is only around 5\,MeV by eight layers of the bars and to secure space for the other detectors.
In addition, using this scintillator bar, it is possible to cover an area of approximately $\SI{1}{\m} \times \SI{1}{\m}$.

\subsubsection{Wavelength-shifting fiber}

The scintillation light from the bar is collected and transported through a wavelength-shifting fiber, Kuraray Y11(200).
The diameter of the fiber is \SI{1}{\mm} and the attenuation length is longer than \SI{3.5}{\m}.
It is placed on the groove of each scintillator bar and glued using an optical cement.

\subsubsection{Multi-pixel photon counter}

The Hamamatsu MPPC is a photon-counting silicon sensor that uses avalanche photo-diodes operated in parallel in the Geiger mode.
One edge of the WLS fiber on each scintillator bar is connected to one MPPC.
In the scintillation tracker, S13081-050CS(X1) MPPCs are used.
Parameters of the MPPC is summarized in Table~\ref{tab:design:mppc_param}.
\begin{table}[t]
    \centering
    \caption{Parameters of MPPC S13081-050CS(X1)~\cite{Yoshida:mt}\label{tab:design:mppc_param}}
    \begin{tabular}{cc} \hline
        Size of a sensitive area & $\SI{1.3}{\mm} \times \SI{1.3}{\mm}$ \\
        Number of pixels & 667 \\
        Size of one pixel & $\SI{50}{\um} \times\SI{50}{\um}$ \\
        Operation voltage & $\sim \SI{54}{\V}$ \\
        Noise rate ($ > 0.5\,\mathrm{p.e.}$, \SI{25}{\degreeCelsius}) & $< \SI{100}{\kHz}$ \\
        Cross-talk rate & $\sim 1\%$ \\
        Photon detection efficiency & $\sim 35\%$ \\
        Sensitive wavelength & 320--\SI{900}{\nm} \\
        Most sensitive wavelength &  \SI{460}{\nm}\\ \hline
    \end{tabular}
\end{table}

\begin{figure}[h]
\centering
\includegraphics[width = 0.6\textwidth]{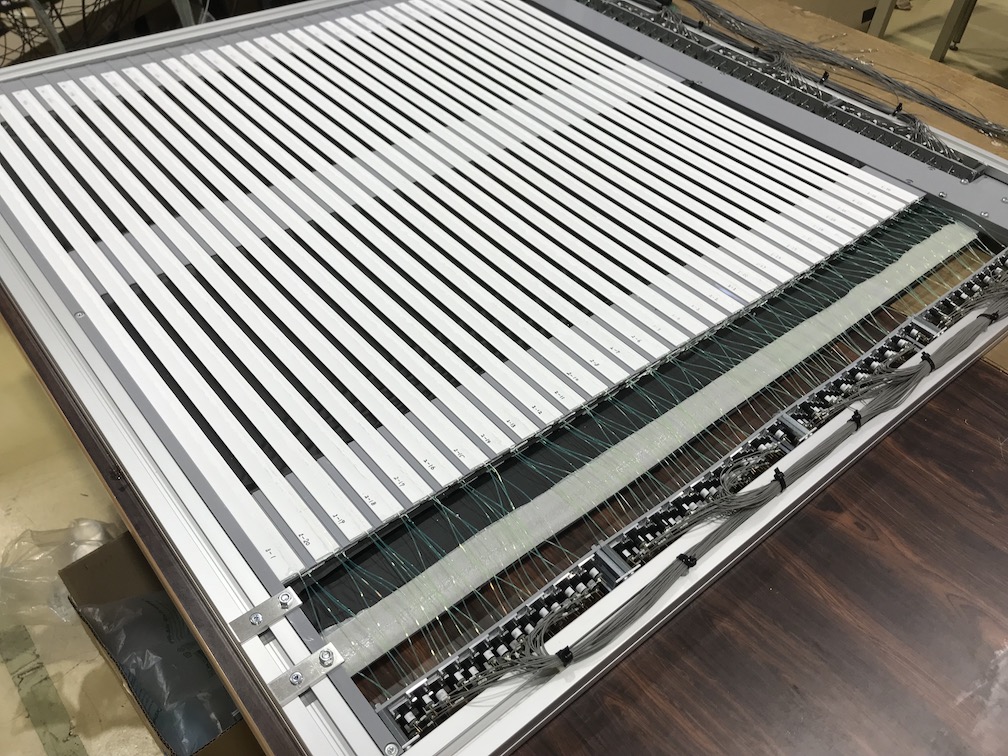}
\caption{The scintillation tracker for the NINJA physics run. Scintillation light from each plastic scintillator bar is read out by the wavelength-shifting fiber and the MPPC.\label{fig:design:tracker_photo}}
\end{figure}

Figure~\ref{fig:design:tracker_photo} shows a photograph of the scintillation tracker after the construction.

\subsection{DAQ system}

NIM EASIROC (Extended Analogue Si-pm Integrated ReadOut Chip) modules~\cite{Nakamura:EASIROC} are used to read out signals from the MPPCs.
Each EASIROC module can operate 64 MPPCs at once and thus, four modules are operated in parallel.
Trigger threshold for each channel is set to 2.5\,p.e. (photoelectron equivalent) and channels exceeding the threshold are treated as hits.
The total rate of the dark noise exceeding the trigger threshold is measured to be \SI{300}{\Hz} before the operation.
When there are hits in both the horizontal and vertical modules, and the beam trigger is provided by the J\nobreakdash-PARC neutrino beamline, the DAQ is triggered and the charge and hit timing in each channel are recorded.

If there are no hits in a spill of the beam, a dummy trigger is generated by delaying the beam trigger by \SI{5}{\us}.
The dummy trigger assures that the scintillation tracker acquires data once in each spill even if there are no hits in the scintillation tracker in the spill.

\subsection{Light yield performance}
The mean light yield of the scintillator bar is measured to be more than 10\,p.e. when the minimum ionizing particle penetrates the bar along the beam direction.
Assuming the Poisson distribution, the variation of the light yield is approximately 30\%, and the inefficiency of the single scintillator bar due to the 2.5\,p.e. trigger threshold will be less than 0.3\%
As described in Section~\ref{ssec:design:design}, the scintillation tracker only uses hit or unhit information to reconstruct the position.
The light yield is sufficient to distinguish between hit and unhit.

\section{Operation\label{sec:operation}}

There were two periods of neutrino beam exposure.
The first one is from November 2019 to December 2019 and the second one is from January 2020 to February 2020.
The scintillation tracker was operated during the whole period of the exposure.
The details of the accumulated POT are shown in Table~\ref{tab:operation:summary}.
The data acquisition efficiency is more than 99.9\% and the inefficiency is due to run changes of the DAQ of the scintillation tracker.
\begin{table}[h]
\centering
\caption{Summary of the POT of the NINJA physics run\label{tab:operation:summary}}
\begin{tabular}{lcc} \hline
Period & Recorded ($10^{20}$) & Delivered ($10^{20}$) \\ \hline
11/7/2019 - 12/19/2019 & 2.64841 & 2.64957 \\
1/14/2020 - 2/12/2020 & 2.11531 & 2.11645 \\ \hline
Total & 4.76372 & 4.76602 \\ \hline
\end{tabular}
\end{table}

\begin{figure}[h]
\centering
\includegraphics[width = 0.99\textwidth]{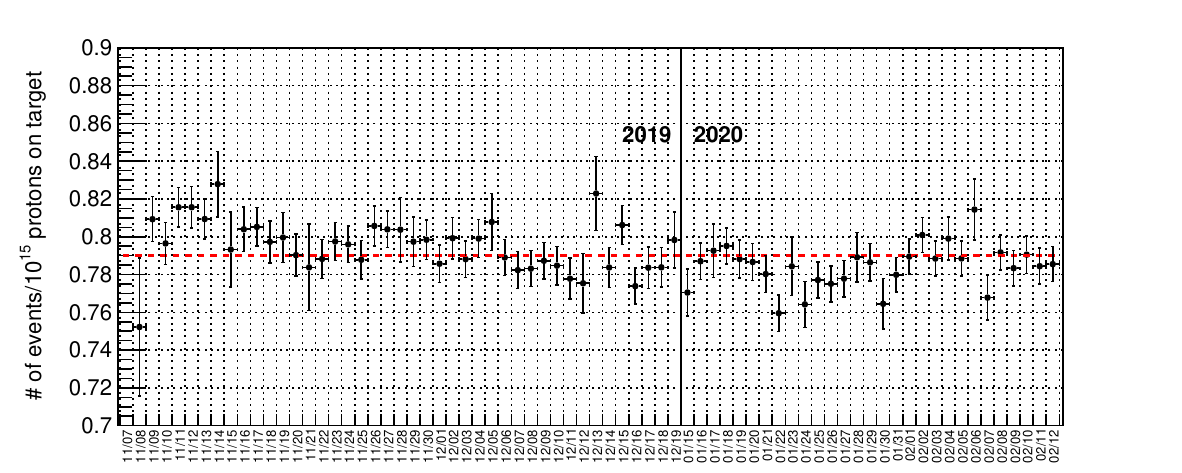}

\caption{Event rate of the tracker. The event rate is stable for the whole period.\label{fig:operation:sandmuon}}
\end{figure}

Figure~\ref{fig:operation:sandmuon} shows the number of observed events per POT in the scintillation tracker in each day.
Here, when more than three hits are observed in one spill, it is defined as one observed event.
The event is dominated by the muons generated by the $\nu_\mu$ CC interactions in the upstream wall of the detector hall.
Therefore the number of observed events in the scintillation tracker is expected to be proportional to the POT.
A stable operation of the scintillation tracker was observed during the whole period of the NINJA physics run.

\section{Muon track matching and position/angle reconstruction\label{sec:matching}}

\begin{figure}[h]
    \centering
    \includegraphics[width = 0.6\textwidth]{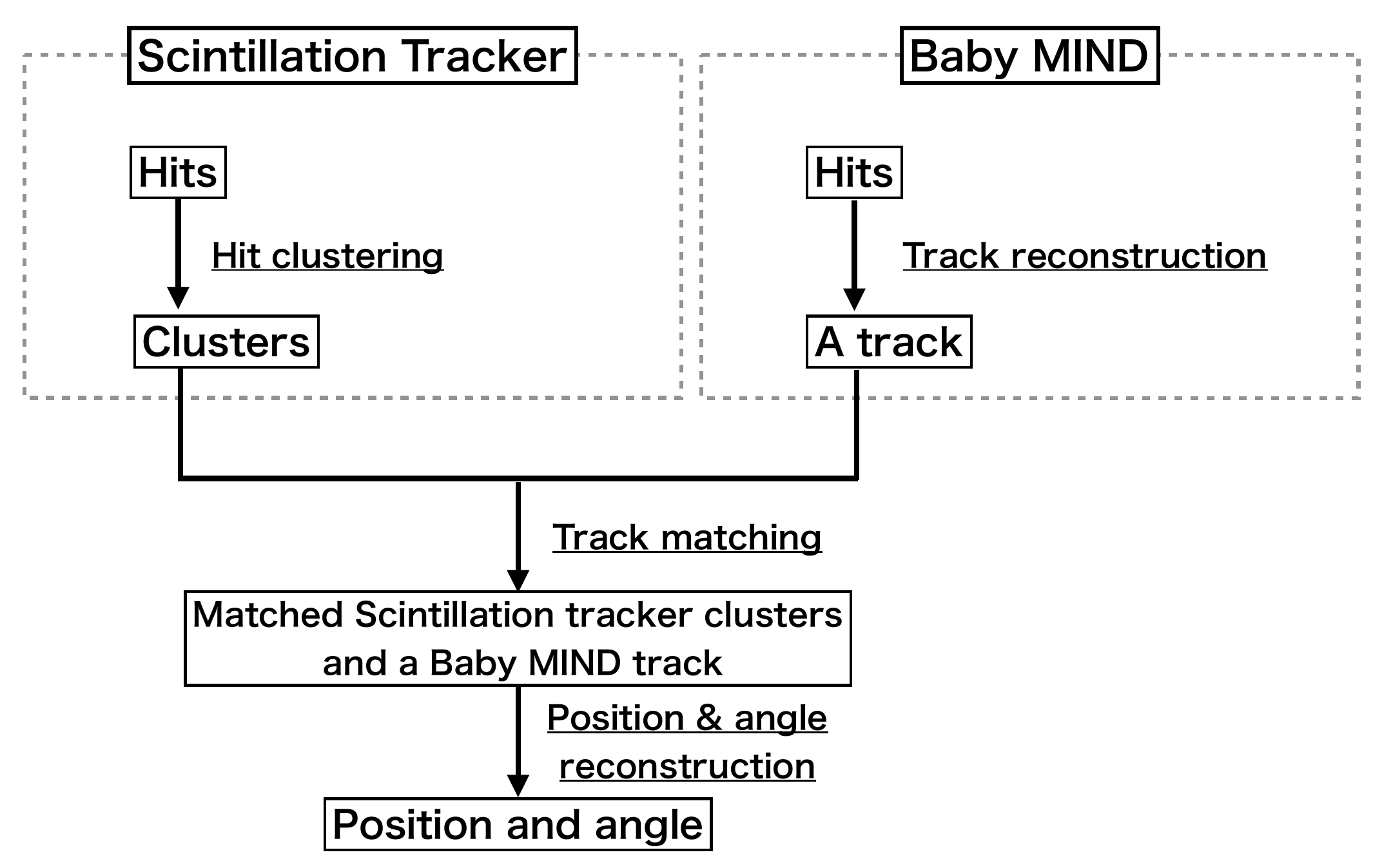}
    \caption{Analysis flow of the hit-track matching and the reconstructions of the position and angle of the muons.\label{fig:matching:flow}}
\end{figure}

In this section, the matching between the hits in the scintillation tracker and the tracks in Baby MIND, and the reconstructions of the position and angle of the muons are described.
The analysis consists of four steps described in Fig.~\ref{fig:matching:flow}; (1) Hit clustering, (2) track reconstruction, (3) hit-track matching, and (4) position and angle reconstructions.

\subsection{Hit clustering in the scintillation tracker}

\begin{figure}
    \centering
    \includegraphics[width = 0.7\textwidth]{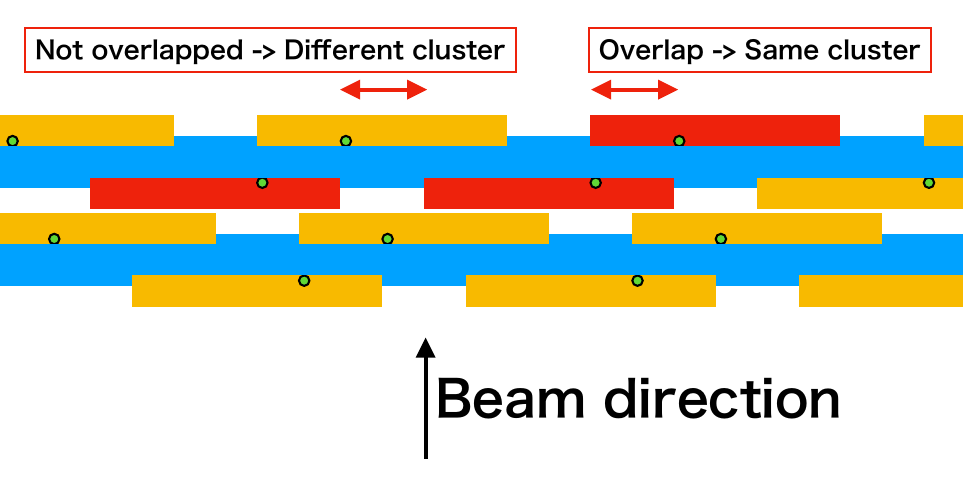}
    \caption{Clustering of the hits in the scintillation tracker. When scintillator bars are overlapped, they are classified into the same cluster; otherwise, they are separated into different clusters.\label{fig:matching:overlap}}
\end{figure}

\begin{figure}[h]
    \centering
    \includegraphics[width = 0.9\textwidth]{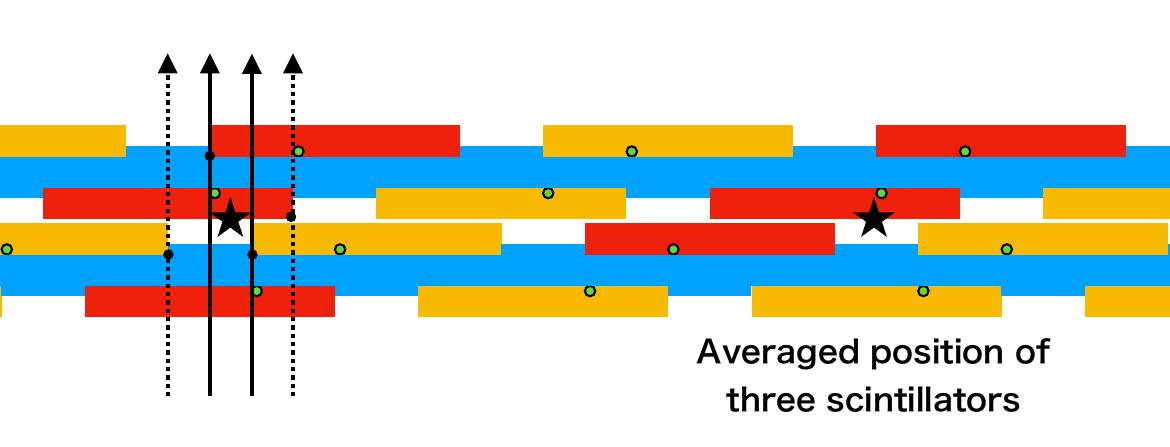}
    \caption{Schematic view of the reconstruction of the tentative position in the scintillation tracker. The red scintillators are the hit scintillators. For the left cluster, black solid lines determine the minimum and maximum position candidates and position is reconstructed as their average. Dotted lines are examples not satisfying the hit pattern. The perpendicular line cannot make the pattern in the right cluster, and the averaged position of all hit scintillators are used instead. Each star represents the reconstructed position of the cluster.\label{fig:matching:pre_posrecon_schematic}}
\end{figure}

The hits in the scintillation tracker are clustered in the horizontal and vertical modules separately.
As shown in Fig.~\ref{fig:matching:overlap}, when scintillator bars with hits are overlapped viewed from the beam direction, they are classified into the same cluster; otherwise, they are separated into different clusters.
The position of each cluster is tentatively reconstructed assuming the particle goes through in a direction perpendicular to the layer of the scintillation tracker.
The hit pattern of the cluster depends on the position of the muon.
The minimum and maximum positions which satisfy the pattern of the cluster are estimated according to the structure of the scintillator bars and their average value is taken as the position of the cluster.
Figure~\ref{fig:matching:pre_posrecon_schematic} shows a schematic of the reconstruction of the tentative position of one cluster in the scintillation tracker.
When the perpendicular line is impossible to make the pattern of the cluster, e.g. the right side of Fig.~\ref{fig:matching:pre_posrecon_schematic}, the average of all hit scintillator positions is used as the position.

\subsection{Track reconstruction in Baby MIND}

The position and angle of the track in Baby MIND are computed by the linear fitting to the hits.
The fitted line is interpolated to the second layer of Baby MIND, and the interpolated point and the slopes are respectively considered as the reconstructed positions and angles of the track in Baby MIND.

\subsection{Hit-track matching between the scintillation tracker and Baby MIND}

\begin{figure}[h]
    \centering
    \includegraphics[width = 0.45\textwidth]{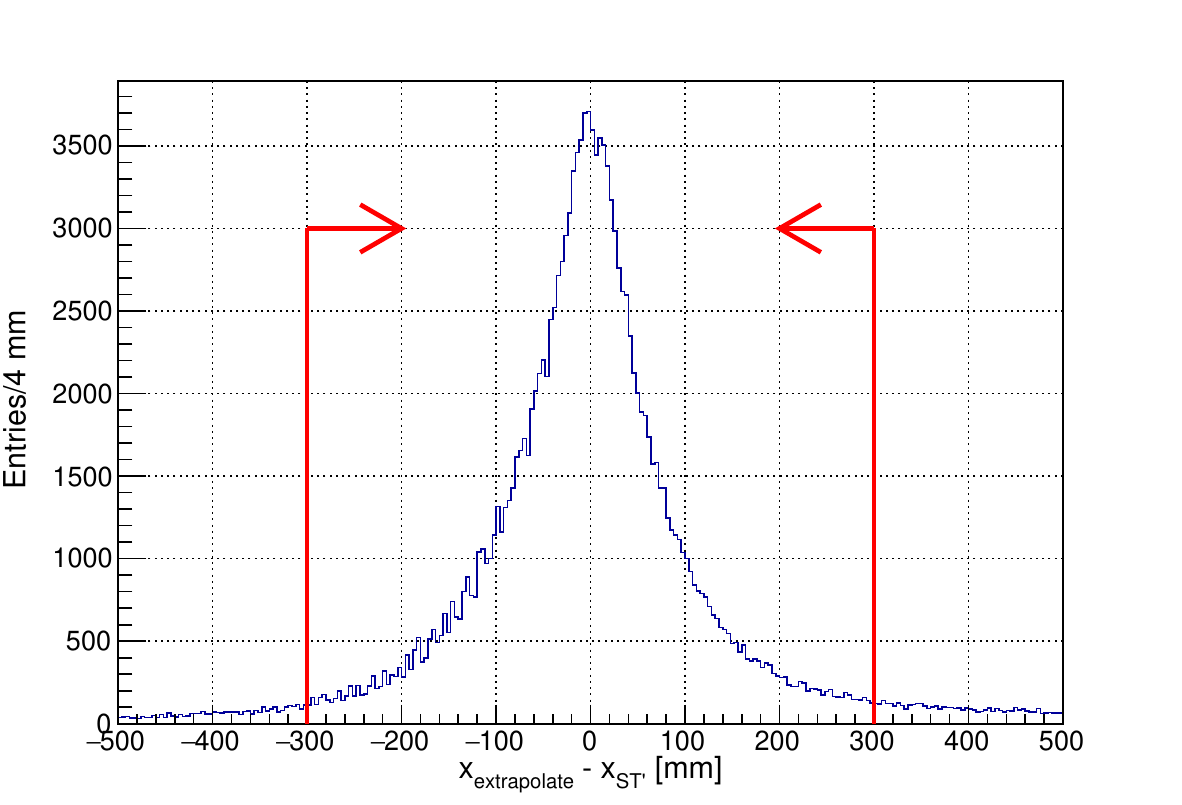}
    \includegraphics[width = 0.45\textwidth]{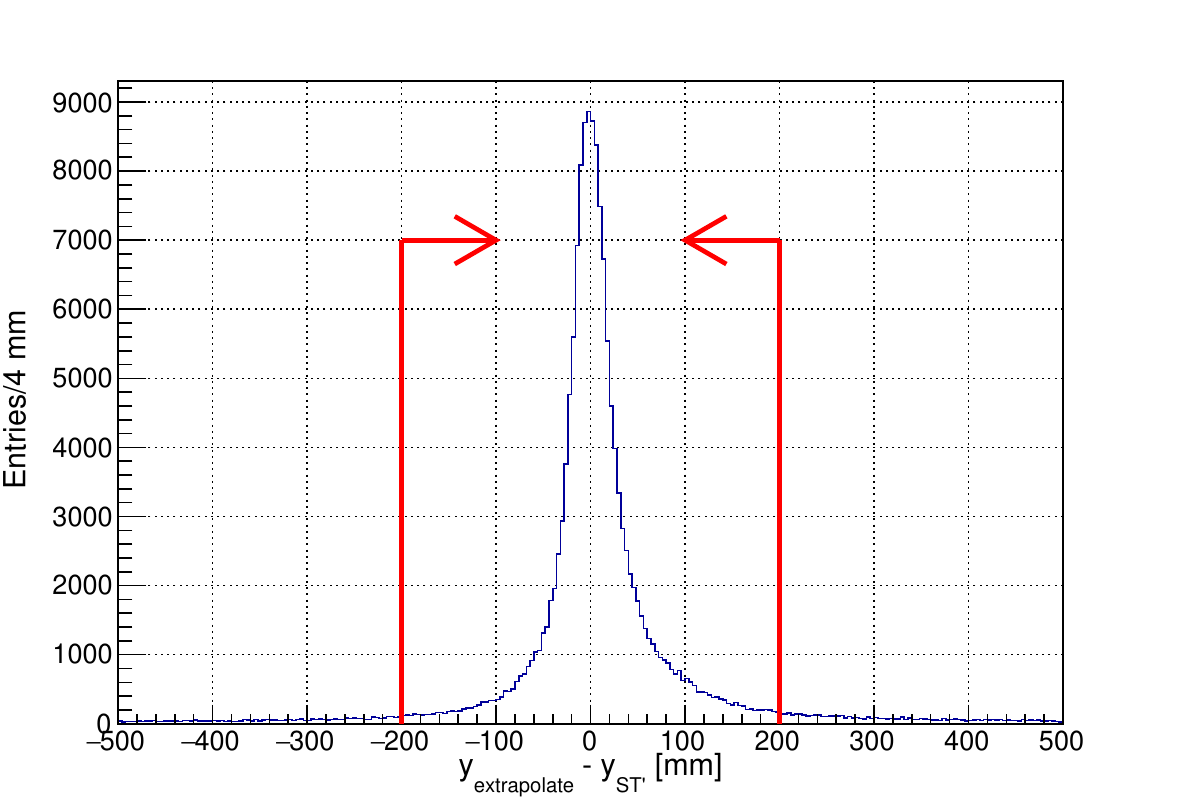}
    \caption{Distributions of the differences between the tentative positions in the scintillation tracker and the extrapolated positions from Baby MIND. The left and right figures show the horizontal and vertical distributions, respectively. The selection criteria are shown with red lines.\label{fig:matching:bmst_match_judge}}
\end{figure}

The tracks in Baby MIND are extrapolated from the second layer ($z \simeq \SI{75}{\cm}$) to the location of the scintillation tracker ($z = \SI{0}{\cm}$).
It is checked whether or not they fall within the sensitive area of the scintillation tracker.
Figure~\ref{fig:matching:bmst_match_judge} shows the differences between the tentative positions in the scintillation tracker shown in Fig.~\ref{fig:matching:pre_posrecon_schematic} and the extrapolated positions from Baby MIND in data.
The left and right figures show the horizontal and vertical distributions, respectively.
The width of the distributions represents the positional and angular resolutions of the tracks in Baby MIND.
The horizontal resolutions are worse than the vertical ones because of the difference in the scintillator width as mentioned in Section~\ref{sssec:setup:requirement:posres}.

In the hit-track matching, the differences between the tentative positions in the scintillation tracker, $x(y)_\mathrm{ST'}$ and the Baby MIND extrapolated position, $x(y)_\mathrm{extrapolate}$ are checked.
Here, the $'$ indicates that the position is tentative.
\begin{align}
	|x_\mathrm{extrapolate} - x_\mathrm{ST'}| &< 300\,\mathrm{mm} \label{eq:bmst_match_allowance_x} \\
	|y_\mathrm{extrapolate} - y_\mathrm{ST'}| &< 200\,\mathrm{mm} \label{eq:bmst_match_allowance_y}.
\end{align}
The cluster with $x(y)_\mathrm{ST'}$ satisfying Eqs.~\eqref{eq:bmst_match_allowance_x} and \eqref{eq:bmst_match_allowance_y} is selected as the matched cluster.
When there are multiple candidates of the clusters to be matched, the candidate with the smallest difference is selected in this analysis.
When the track has both the horizontal and vertical matched clusters, it is considered to be matched with the hits in the scintillation tracker.
The selection criteria are shown as red lines in Fig.~\ref{fig:matching:bmst_match_judge}.

\subsection{Reconstruction of angle\label{ssec:matching:angrecon}}

\begin{figure}[H]
\centering
\includegraphics[width = 0.7\textwidth]{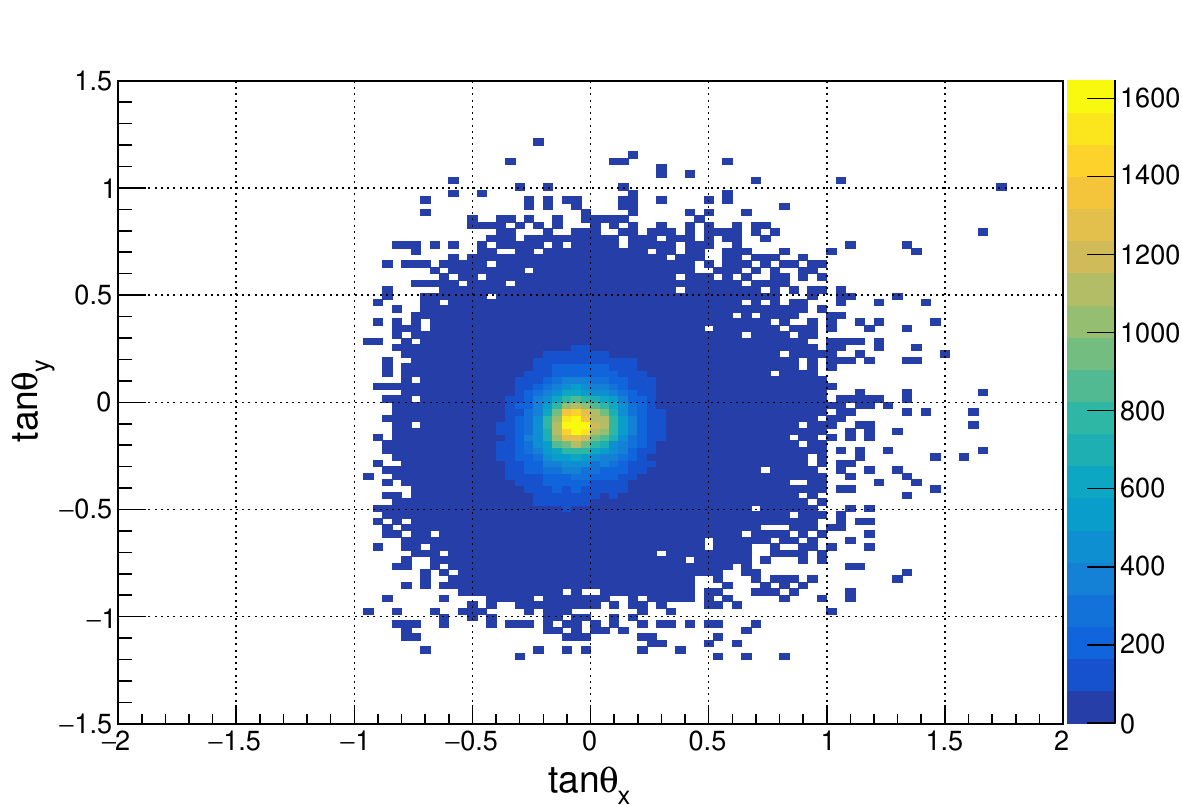}
\caption{Distribution of two-dimensional angles reconstructed from the scintillation tracker and Baby MIND. The layers of the scintillation tracker are almost perpendicular to the neutrino beam and the distribution is consistent with it.\label{fig:matching:ntbmangrecon_2d}}
\end{figure}

The basic concept of the angle reconstruction was described in Section~\ref{sec:setup}.
After the hit-track matching, the tentative position in the scintillation tracker is used as $x(y)_\mathrm{ST}$, and the position or the track at the second layer in Baby MIND ($z \simeq \SI{75}{\cm}$) is used as $x(y)_\mathrm{BM}$ in Eq.~\eqref{eq:setup:thetarec}, respectively.
The distribution of the reconstructed angles ($\tan\theta_{x(y)}$) is shown in Fig.~\ref{fig:matching:ntbmangrecon_2d}.
The distribution shows a peak around $\tan\theta = 0$.
The layers of the scintillation tracker are almost perpendicular to the neutrino beam and the distribution is consistent with it.
A small shift from $\tan\theta = 0$ is due to the difference in the position of the center of the neutrino beam and the position of the scintillation tracker.

\subsection{Reconstruction of position\label{ssec:matching:posrecon}}

\begin{figure}[h]
    \centering
    \includegraphics[width = 0.5\textwidth]{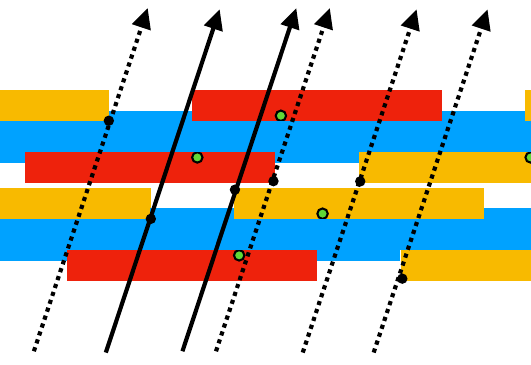}
    \caption{Schematic view of the reconstruction of the position in the tracker. The red scintillators are the hit scintillators in one cluster. Black solid lines determine the minimum and maximum position candidates and dotted lines are examples not satisfying the hit pattern.\label{fig:matching:posrecon_schematic}}
\end{figure}

The final position reconstruction of the scintillation tracker-Baby MIND matched tracks is almost the same as the reconstruction of the tentative position in the scintillation tracker.
The only difference is that the particle penetrates the layer not necessarily in a perpendicular direction.
Figure~\ref{fig:matching:posrecon_schematic} shows the lines that have the reconstructed angle as the slope and extrapolated from each scintillator vertex.
They are checked if it can make the pattern of the hit cluster.

\begin{figure}[H]
\centering
\includegraphics[width = 0.7\textwidth]{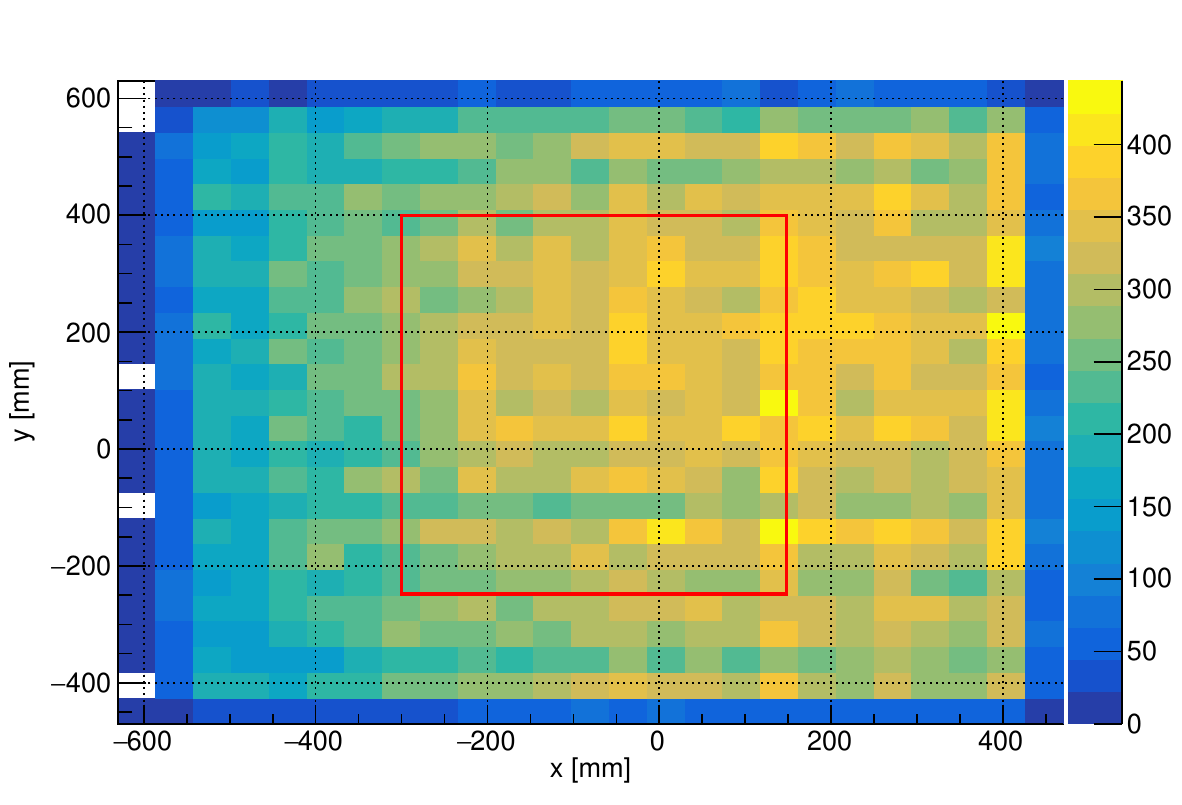}
\caption{Positional distribution reconstructed from the scintillation tracker and Baby MIND. The distribution is not uniform since the scintillation tracker and Baby MIND are not centered and the angular acceptances for the muon tracks are different for different positions. A red rectangle shows an area used in the evaluation of the efficiency described in Section~\ref{ssec:performance:matcheff}.\label{fig:matching:ntbmposrecon_2d}}
\end{figure}

The distribution of the reconstructed positions is shown in Fig.~\ref{fig:matching:ntbmposrecon_2d}.
The distribution is not uniform since the central positions of the tracker and Baby MIND are not identical; hence, the angle acceptance for the muon tracks depends on the positions.

\section{Performance of the track matching and reconstruction\label{sec:performance}}

\subsection{Efficiencies\label{ssec:performance:matcheff}}

Efficiency is evaluated using the data obtained during the operation of the scintillation tracker and Baby MIND.
The efficiency is defined as a ratio of the number of the matched tracks to the number of the tracks in Baby MIND.
By definition, this efficiency is a product of the detection efficiency of the scintillation tracker and the hit-track matching efficiency between the scintillation tracker and Baby MIND.
To ensure that the muons are not generated between the emulsion shifter and the scintillation tracker, but are coming from the upstream of the detectors, the tracks are required to have hits in the upstream WAGASCI or Proton Module in Fig~\ref{fig:setup:wagasci_babymind_w_ninja_w_note}.
To ensure the tracks penetrate the sensitive area, the tracks are required to be in the inner area ($\SI{400}{\mm} \times \SI{600}{\mm}$) at the position of the scintillation tracker shown as a red rectangle in Fig.~\ref{fig:matching:ntbmposrecon_2d}.

\begin{figure}[h]
\centering
\includegraphics[width = 0.7\textwidth]{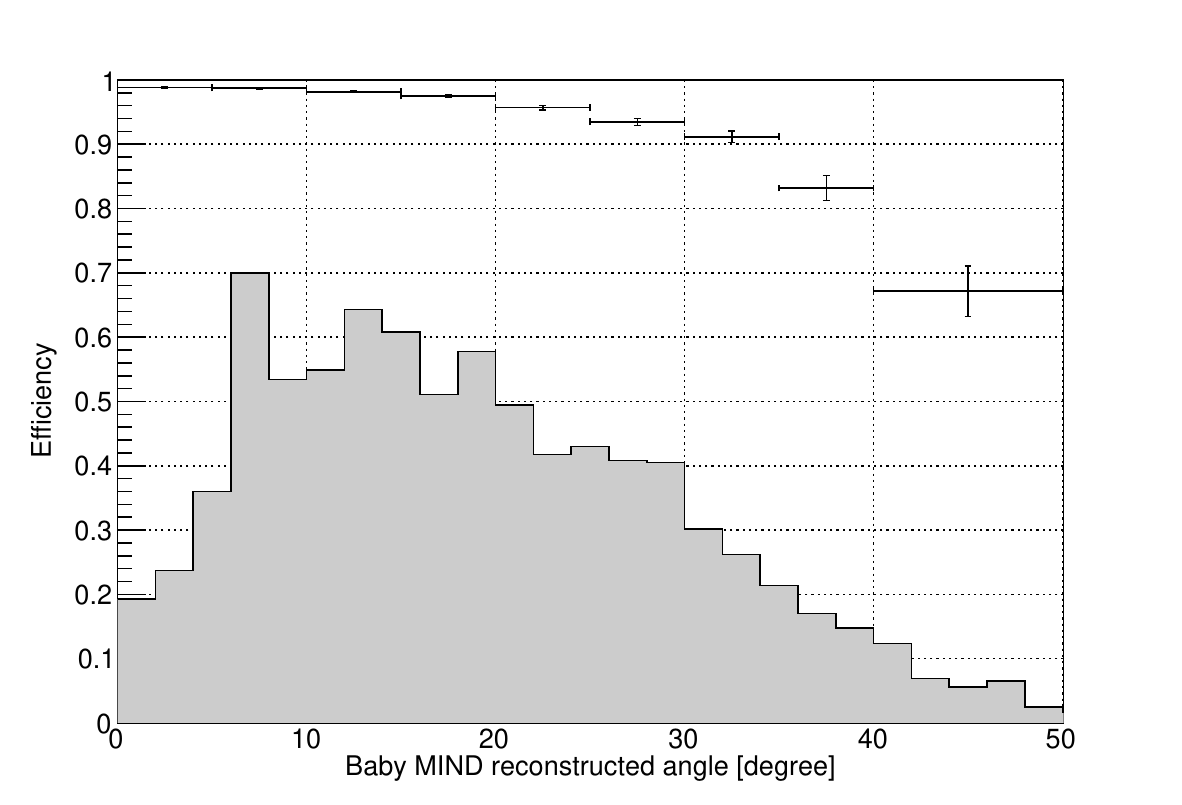}
\caption{Product of the detection efficiency of the scintillation tracker and the matching efficiency between the scintillation tracker and Baby MIND. Expected distribution of Baby MIND reconstructed angle of muon tracks from neutrino-water interactions in the ECCs is shown by a gray histogram.\label{fig:matching:matcheff}}
\end{figure}

Figure~\ref{fig:matching:matcheff} shows the efficiency.
The histogram shows the expected angular distribution of the muon tracks reconstructed by Baby MIND from neutrino-water interactions in the ECCs calculated by the Monte Carlo simulation.
In the forward region ($\theta < \ang{25}$), the efficiency is 97--98\% as shown in Fig.~\ref{fig:matching:matcheff}.
The averaged value of the product of the detection efficiency of the scintillation tracker and the matching efficiency between the scintillation tracker and Baby MIND for the expected muons from the neutrino interactions in the ECCs is 95\%, which achieves the target value described in Section~\ref{sssec:setup:requirement:efficiency}.
In the large angle region, the angle of the low-momentum or large-angle muon tracks tends to be poorly reconstructed because of scattering and small number of hits in Baby MIND.
As a result, the track matching efficiency deteriorates in that region.
On the other hand, the detection efficiency of the scintillation tracker is expected to be nearly 100\% even for the large angle region.
In the future analysis of the $\nu_\mu$ CC interactions, the track matching criteria will be optimized as a function of angle and momentum, and the efficiency will be improved.

\subsection{Positional and angular resolutions for the reconstructed muon tracks\label{ssec:performance:resolution}}

The tracks matched between the scintillation tracker and Baby MIND are then matched with the emulsion shifter.
A collection of tracks in the emulsion shifter in four hours is selected by the timing information of the track reconstructed by the scintillation tracker and Baby MIND.
Then a track which has the smallest positional and angular differences is matched to scintillator detectors.
The detailed information of the matching will be described in another paper.
The positional and angular resolutions are evaluated using the muon tracks matched among the emulsion shifter, the scintillation tracker, and Baby MIND.
Since the positional and angular resolutions of the emulsion shifter are much better than those of the scintillation tracker, the variances in the positional or angular differences between the emulsion shifter and the scintillation tracker-Baby MIND represent the resolutions of the matched tracks between the scintillation tracker and Baby MIND.

\begin{figure}[h]
\centering
\includegraphics[width = 0.45\textwidth]{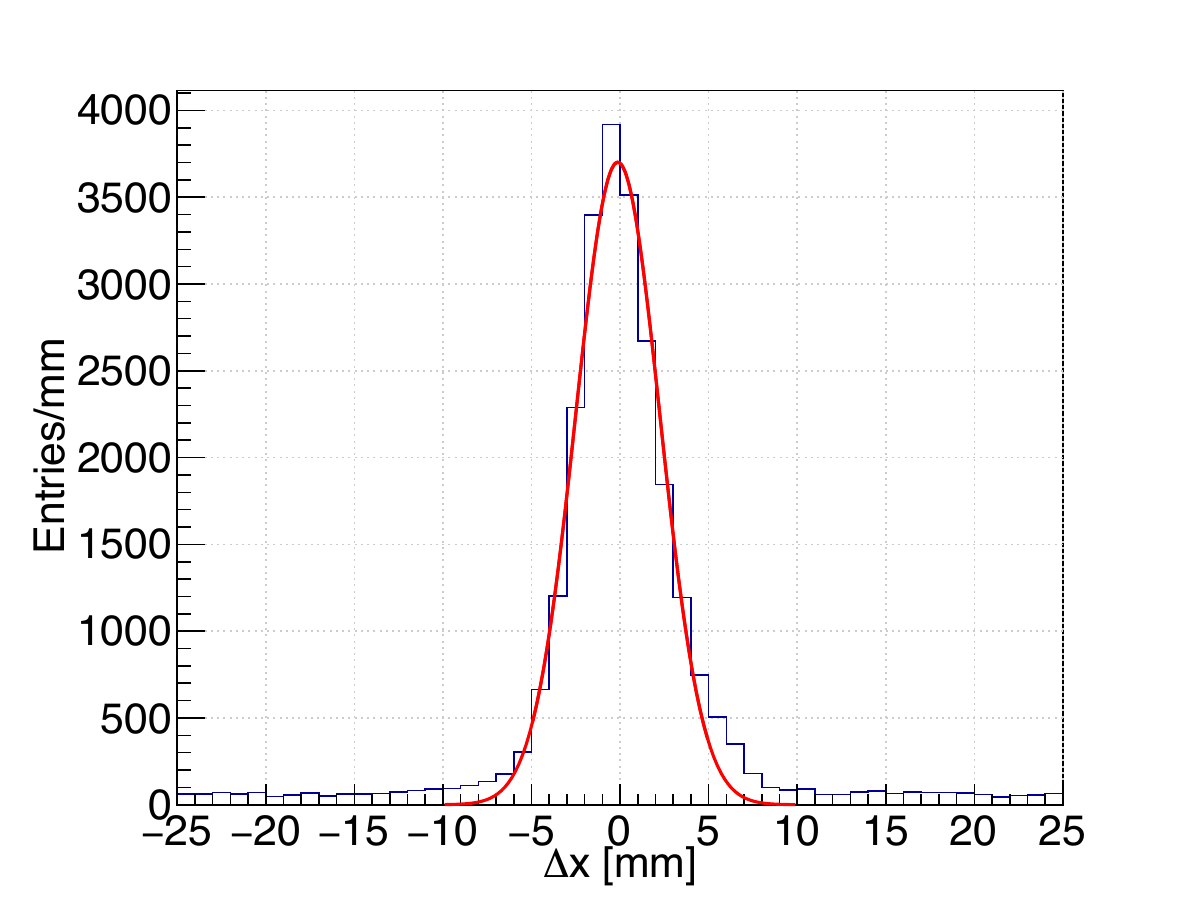}
\includegraphics[width = 0.45\textwidth]{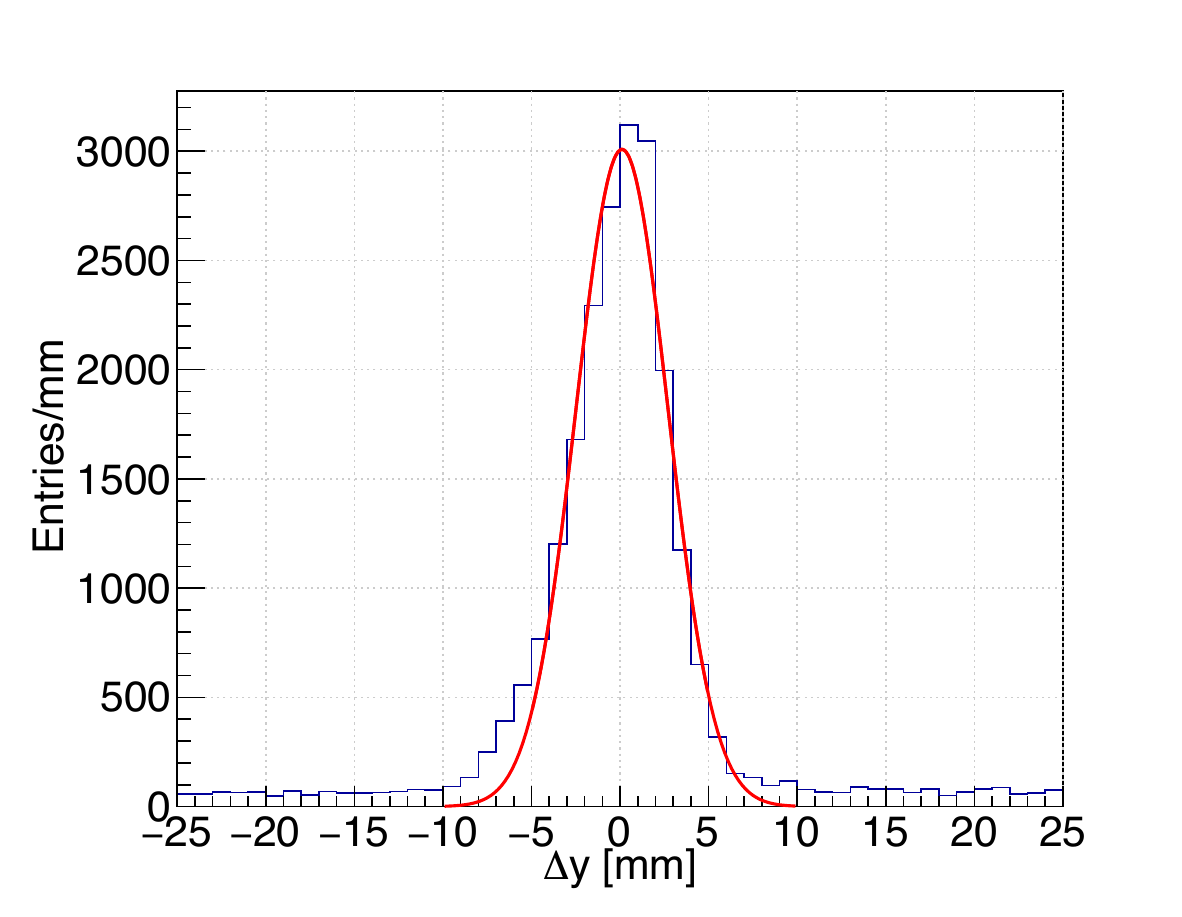}
\caption{Positional difference distributions of the scintillation tracker and the emulsion shifter. The left shows the vertical and the right shows the horizontal distributions.\label{fig:resolution:posres}}
\end{figure}

Figure~\ref{fig:resolution:posres} shows the distribution of the positional differences.
The left and right figures show the horizontal ($\Delta x$) and vertical ($\Delta y$) distributions, respectively.
The distributions are fitted by the Gaussian functions and the width $\sigma \sim \SI{2.5}{\mm}$ satisfies the requirement described in Section~\ref{sssec:setup:requirement:posres}.
The offset of the distributions are attributed to the chance coincidence with cosmic muons in the emulsion shifter, and the tails around $|\Delta x(y)| \lesssim \SI{10}{\mm}$ outside the fitted Gaussian function are due to low-momentum muons.
The low-momentum muons tend to scatter with large angles and the positional differences increase.
The performance of the scintillation tracker is evaluated using the fitted Gaussian function because the component inside the function is mainly considered to be comprising high-momentum muons.
They are less affected by the scattering and the positional resolution can be evaluated correctly.

\begin{figure}[H]
\centering
\includegraphics[width = 0.7\textwidth]{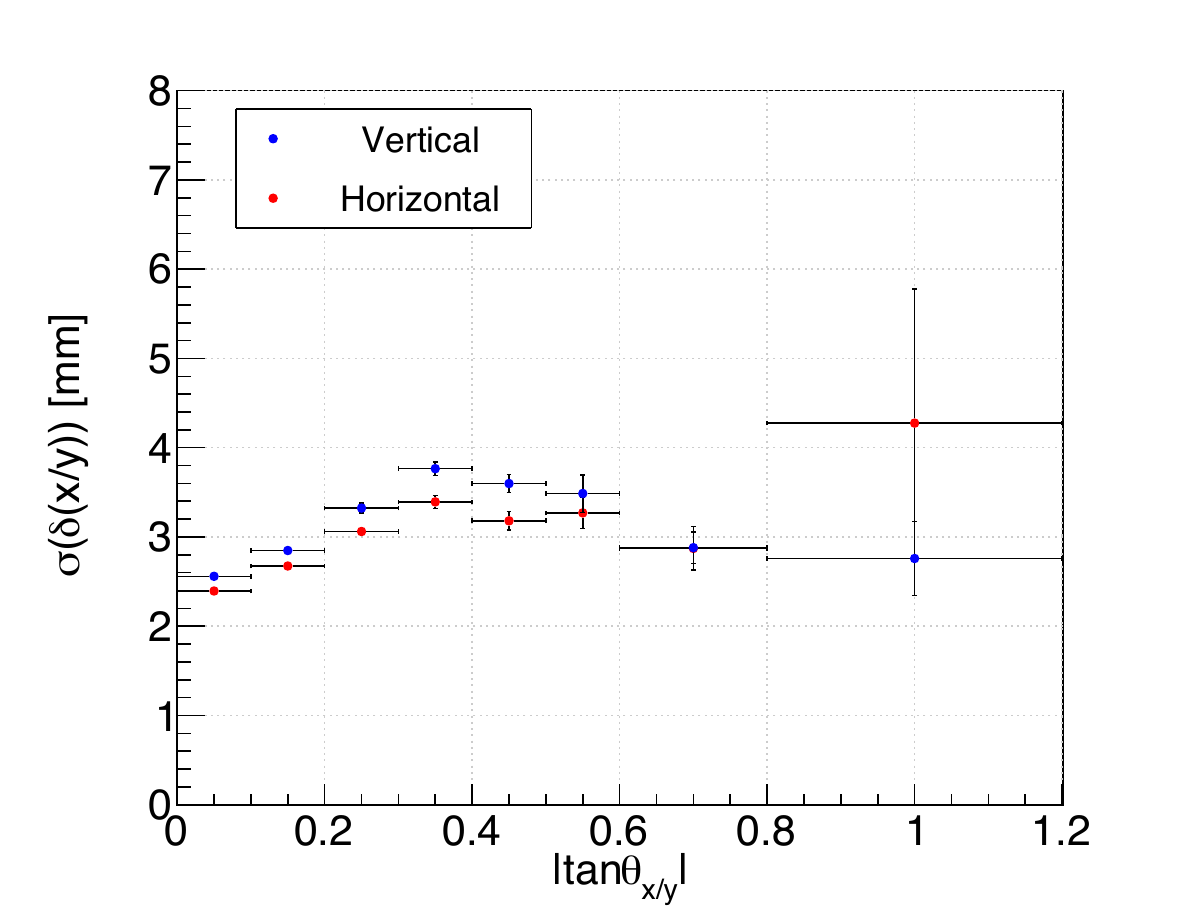}
\caption{Angular dependence of the positional resolution of the scintillation tracker. The points are obtained from the Gaussian fits to each distribution, and the vertical errors are statistical ones obtained from the fittings. The histogram is expected distribution of the angle of muons from the neutrino-water interactions in the ECCs.}\label{fig:resolution:posres_tangent}
\end{figure}

Figure~\ref{fig:resolution:posres_tangent} shows the obtained $\sigma$ values for the horizontal and vertical directions as a function of the track angle.
For forward-directing tracks, the positional resolution of the scintillation tracker is around \SI{2.5}{\mm} and it is less than \SI{3.5}{\mm} even in the large angle region ($|\tan\theta_{x(y)}| \gtrsim 0.5$).
As the horizontal and vertical modules are identical to each other and the reconstruction of the position is primarily performed using only the information of the scintillation tracker, their resolutions are similar as expected.
The histogram shows the expected distribution of $\tan\theta$ of muons from the neutrino interactions calculated by the Monte Carlo simulation.

\begin{figure}[h]
\centering
\includegraphics[width = 0.45\textwidth]{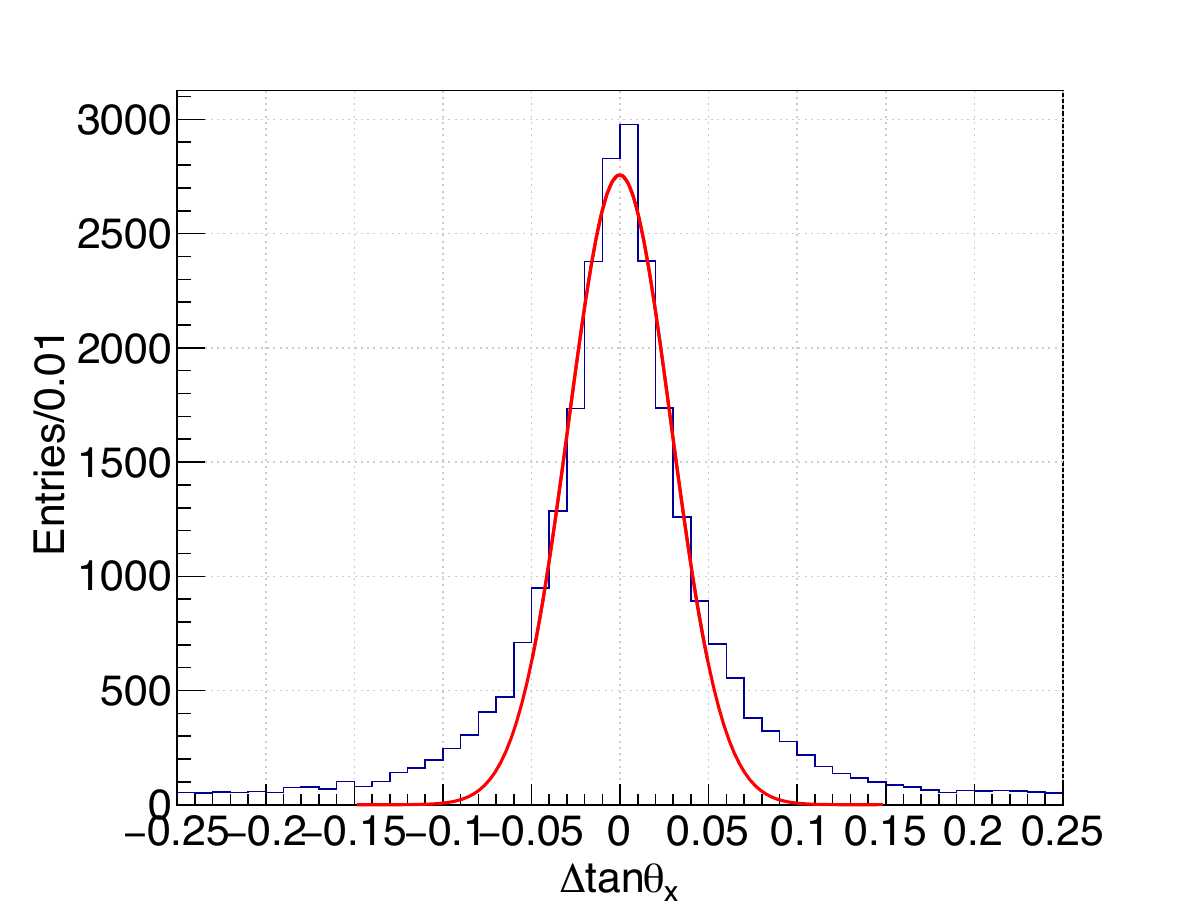}
\includegraphics[width = 0.45\textwidth]{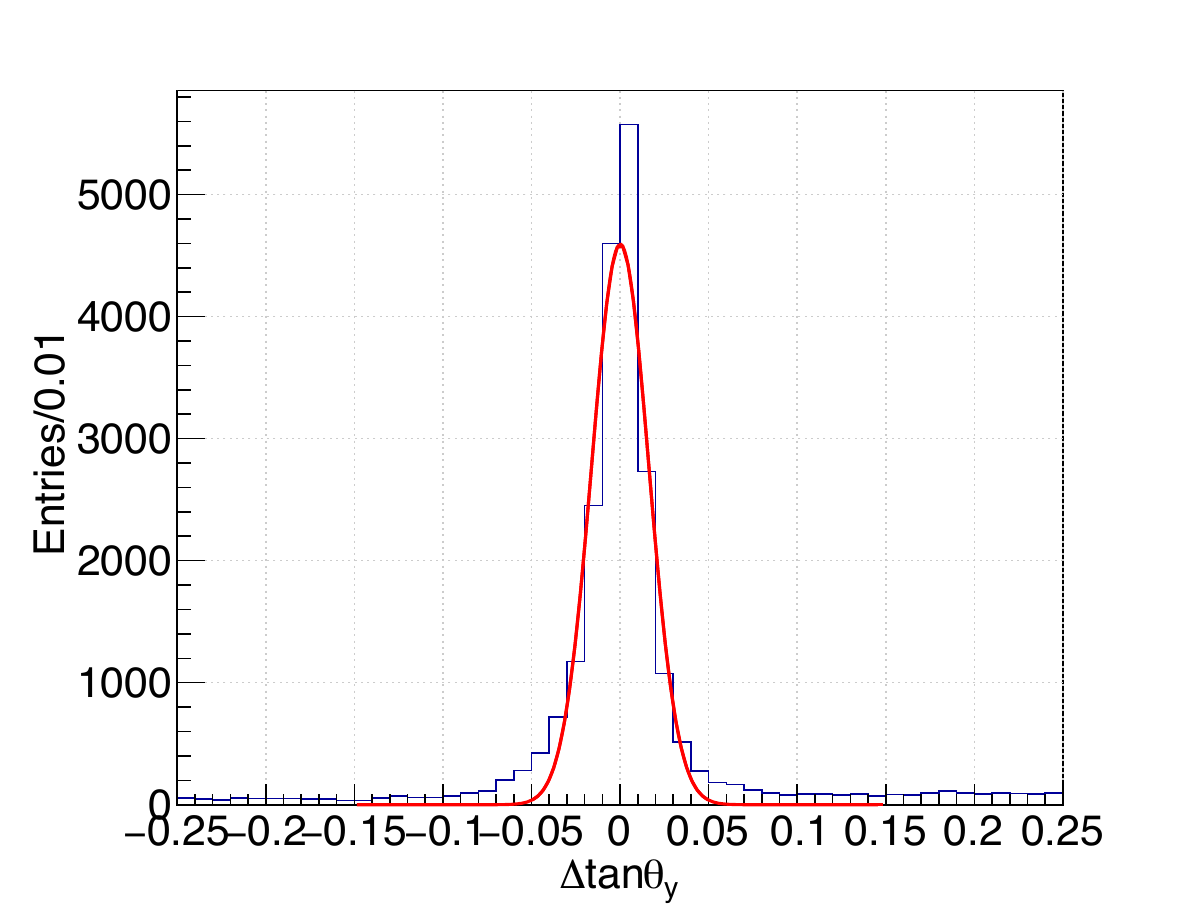}
\caption{Angular difference distribution of the scintillation tracker and the emulsion shifter. The left and right figures show the horizontal and vertical distributions. \label{fig:resolution:angres}}
\end{figure}

The angular resolution is also evaluated using the same tracks as used in the evaluation of the positional resolution.
Figure~\ref{fig:resolution:angres} shows the distribution of the angular differences.
The left and right figures show the horizontal ($\Delta\tan\theta_x$) and vertical ($\Delta\tan\theta_y$) distributions, respectively.
The distributions are also fitted by the Gaussian functions and the width of the horizontal distribution is $\sigma \sim \SI{40}{\milli\radian}$ while that for the vertical one is $\sigma \sim \SI{20}{\milli\radian}$.
It is evident that the angular resolution is different between the horizontal and vertical directions since the scintillator width of Baby MIND is different from each other.
The offset and tail components are attributed to the chance coincidence and the low-momentum muons, respectively, as in the positional difference distributions.

\begin{figure}[h]
\centering
\includegraphics[width = 0.7\textwidth]{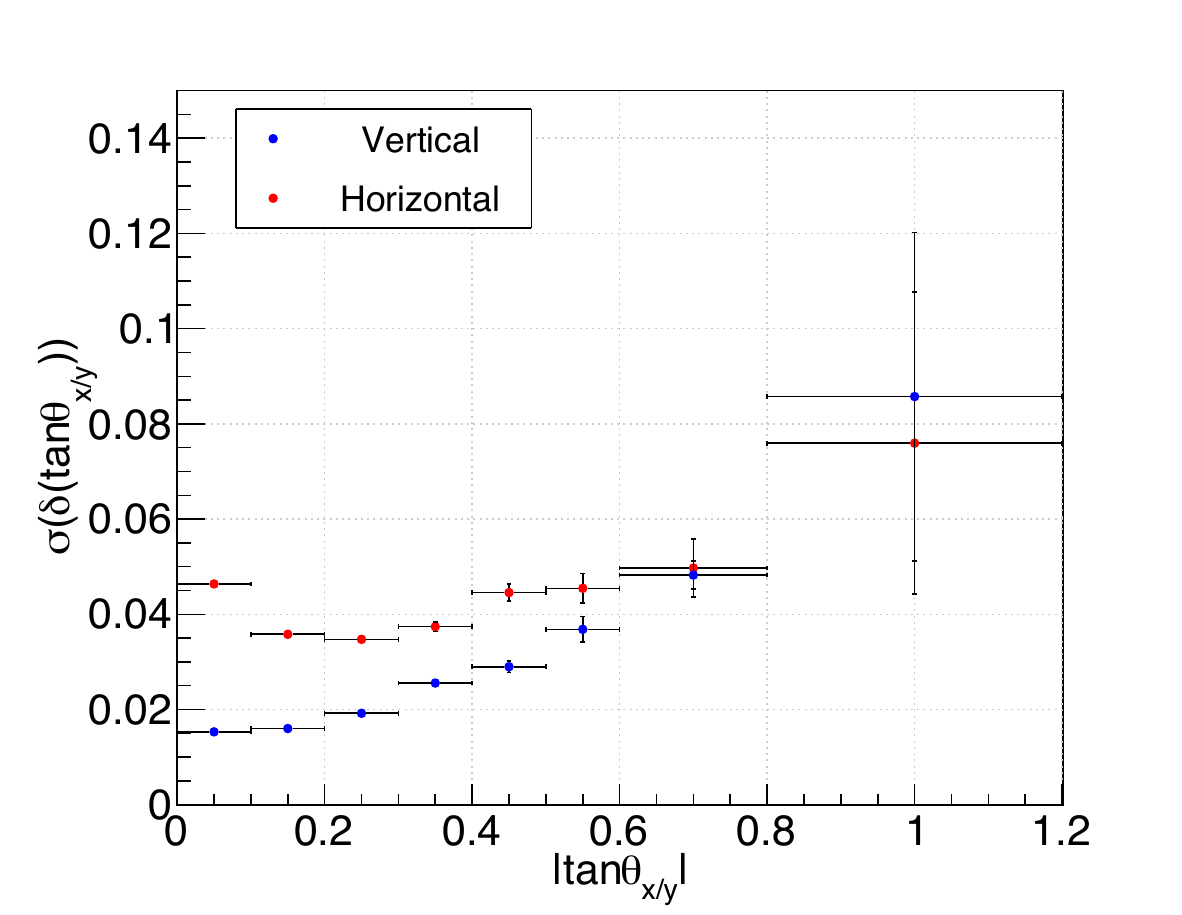}
\caption{Angular dependence of the angular resolution of the scintillation tracker. The points are obtained from the Gaussian fits to each distribution, and the vertical errors are statistical ones obtained from the fittings. The histogram is expected distribution of the angle of muons from the neutrino-water interactions in the ECCs.}\label{fig:resolution:angres_tangent}
\end{figure}

Figure~\ref{fig:resolution:angres_tangent} shows the obtained $\sigma$ values for the horizontal and vertical directions as a function of the track angle.
The angular resolution of muon tracks reconstructed by the scintillation tracker and Baby MIND is less than 100\,mrad for a wide angle range.
The vertical resolution is almost consistent to the expected value ($\sigma \simeq \SI{20}{\milli\radian}$) which is dominated by the scattering as described in Section~\ref{sssec:setup:requirement:posres}.
On the other hand, the horizontal one is twice better than the expected ($\sigma \simeq \SI{81}{\milli\radian}$), since $x_\mathrm{BM}$ is reconstructed by not only one layer information but the linear fitting to the hits.

\subsection{Summary of the performance of the scintillation tracker\label{ssec:peformance:summary}}

The performance of the scintillation tracker satisfies the requirements described in Section~\ref{ssec:setup:requirement}.
The scintillation tracker has more than 97\% efficiency for the forward-directing muon tracks.
The positional resolution is \SI{2.5}{\mm} for both the horizontal and vertical directions.
The angular resolution for the horizontal and vertical directions are \SI{40}{\milli\radian} and \SI{20}{\milli\radian}, respectively.
The resolutions are not consistent to the values discussed in Section~\ref{sssec:setup:requirement:posres} since the expected values are obtained before the track reconstruction in Baby MIND is matured.
The actual positional resolution of Baby MIND has been improved by the linear fitting the hits.
In particular, the horizontal angular resolution is affected by this improvement since the expected value is dominated by the scintillator width of Baby MIND.
The positional resolution of the matched muon tracks is, on the other hand, worse than the ideally expected value, $\SI{24}{\mm} / 6 / \sqrt{12} = \SI{1.2}{\mm}$.
The scintillator bars are actually not straight along the horizontal or vertical direction but slightly sagged around a few mm over \SI{1}{\m}.
Such a sag is not considered in the reconstruction of the position and the positional resolution deteriorates by it.
In particular, the vertical direction is more affected by the gravity and the sag will be larger.
Moreover, mis-alignment of the scintillator bars from the expected position makes the resolution worse.
The effect from the mis-alignment will be also larger in the horizontal direction since the bars tend to be more moved downward by the gravity.
When the detector alignment is adjusted precisely and locally, such effects will be reduced and the positional resolutions will get better.

\section{Summary\label{sec:summary}}

In this paper, the design and performance of the scintillation tracker for the NINJA physics run are described.
The scintillation tracker connects muon tracks between the emulsion detectors and the muon range detector, Baby MIND.
The scintillation tracker covers an area of $\SI{1}{\m} \times \SI{1}{\m}$ which is ten times larger than the previous NINJA runs, while maintaining the number of readout channels less than 256.
The scintillation tracker has a special arrangement of \SI{24}{\mm}-width plastic scintillator bars with deliberate gaps between each other, and the positional resolution is improved to be \SI{2.5}{\mm} using the constraint by the gaps.
The design also maintains the high efficiency of the scintillation tracker since there are no insensitive areas.

The hit-track matching efficiency is evaluated using muon tracks from the upstream of the detectors and an efficiency more than 97\% is achieved for forward-going muons.
The positional and angular resolutions of the scintillation tracker for forward-going muons are estimated to be \SI{2.5}{\mm} and 20-40\,mrad, respectively, using the emulsion shifter information.

It is concluded that the performance of the scintillation tracker is sufficient for the muon track matching in the nuclear-emulsion-based neutrino interaction measurement, which is essential in the $\nu_\mu$ CC interaction analysis of the current and future neutrino experiments using nuclear emulsion.

\section*{Acknowledgment}

The data used in this paper were acquired with the help of the collaborators of the NINJA experiment.
We also received great help from Y.~Ashida and S.~Kuribayashi for the construction and installation.
We thank the T2K collaboration for their support in conducting the NINJA experiment as well as the J-PARC staff for their superb accelerator performance.
We would also like to show our gratitude to the T2K INGRID and WAGASCI/Baby MIND groups for the stable operation, providing experimental data, and cooperation in the software development.
We acknowledge the T2K neutrino beam group for providing a high-quality and stable neutrino beam as well as supporting the neutrino beam MC simulation.
We would like to thank Editage (\url{www.editage.com}) for English language editing.

This work was supported by MEXT and JSPS KAKENHI Grant Numbers JP17H02888, JP18H03701, JP18H05535, JP18H05537, JP18H05541, JP20J15496, and JP20J20304.

\bibliography{main}

\end{document}